\documentclass[a4paper,fleqn,usenatbib]{mnras}

\usepackage{newtxtext,newtxmath,pdflscape}

\usepackage[T1]{fontenc}
\usepackage{ae,aecompl}

\usepackage{graphicx}	
\usepackage{amsmath}	
\usepackage{amssymb}	
\usepackage{multirow}
\usepackage{booktabs}

\hypersetup{draft}

\newcommand*\VS{\color{black}}

\newcommand*\AZ{\color{black}}
\newcommand*\VST{\color{black}}
\newcommand*{\newtwo}{\color{black}}
\newcommand*{\new}{\color{black}}

\title[Multidimensional AGN Classification]{Multidimensional Data Driven Classification of \newtwo{Emission-line} Galaxies }

\author[V. Stampoulis et al.]{
Vasileios Stampoulis,$^{1}$\thanks{E-mail: vs2712@ic.ac.uk }
David A. van Dyk,$^{1}$
Vinay L. Kashyap$^{2}$
and Andreas Zezas$^{2,3,4}$\thanks{E-mail: azezas@physics.uoc.gr }
\\
$^{1}$Statistics Section, Imperial College London, Huxley Building, South Kensington Campus, London SW7, UK\\
$^{2}$Harvard-Smithsonian Center for Astrophysics, 60 Garden St.,
Cambridge, MA 02138, USA\\
$^{3}$Physics Department,  Institute of Theoretical $\&$ Computational Physics,  University of Crete, Heraklion 71003, Greece\\
$^{4}$Foundation for Research and Technology-Hellas,  Heraklion 71110,  Greece
}
\date{Accepted XXX. Received YYY; in original form ZZZ}

\pubyear{2017}

\begin{document}
\label{firstpage}
\pagerange{\pageref{firstpage}--\pageref{lastpage}}
\maketitle

\begin{abstract}
We propose  a  new soft clustering scheme {\VS for classifying galaxies in different activity classes} using simultaneously 4 emission-line ratios; $\log($[\ion{N}{II}]$/$H$\alpha$), $\log$([\ion{S}{II}]$/$H$\alpha$),  $\log$([\ion{O}{I}]$/$H$\alpha$) and $\log$([\ion{O}{III}]$/$H$\beta$). We fit {\VS 20} multivariate Gaussian distributions {\new{to the 4-dimensional distribution of these lines}} obtained from the  Sloan Digital Sky Survey (SDSS)  in order to capture local structures and subsequently group the multivariate Gaussian distributions {\VS to represent the complex multi-dimensional structure of the joint distribution of galaxy spectra in the 4 dimensional line ratio space. } {\VS The main advantages of this method are the use of all four optical-line ratios simultaneously and the adoption of a clustering scheme. This maximises the use of the available information,  avoids contradicting classifications, and treats each class as a distribution resulting in soft classification boundaries {\new{and providing the probability for an object to belong to each class}}.} We also introduce linear multi-dimensional decision surfaces using support vector machines {\VS based on the classification of our soft clustering scheme.  This linear multi-dimensional hard clustering technique shows  high classification accuracy  with respect to our soft-clustering scheme.}
\end{abstract}




\section{Introduction}

{\newtwo The production of electromagnetic radiation} in galaxies is dominated by two main processes: star-formation and/or accretion onto a supermassive central black-hole, the latter witnessed as an Active Galactic Nucleus (AGN). The {\new characterization of  these processes and the study of their interplay} is key for understanding the demographics of galactic activity and the  co-evolution of nuclear black-holes and their host galaxies  \citep[e.g.][]{2013ARAA..51..511K}.  {\newtwo{One of the most commonly used tools}} for characterising the type of activity in galaxies is  its imprint on the emerging spectrum of the photoionised interstellar medium (ISM). AGN generally produce harder ionising continua which result in spectra with stronger high-excitation lines {\new{ compared to the spectra we can obtain from photoionization by young stellar populations}}  \citep[e.g.][]{2003ARAA..41..517F}.
 
 The importance of characterising the ionising source of emission-line regions was recognised early on and led to the first systematic presentation of optical emission-line diagnostic tools by \citet*{baldwin1981classification}. This work introduced two-dimensional diagrams  involving the ratios of various optical emission lines (e.g. $[\ion{Ne}{V}]$\,$\mathrm{\lambda3426\AA}$, $[\ion{O}{II}]$\,$\mathrm{\lambda3727\AA}$, $[\ion{O}{III}]$\,$\mathrm{\lambda5007\AA}$, $[\ion{O}{I}]$\,$\mathrm{\lambda6300\AA}$, $[\ion{N}{II}]$\,$\mathrm{\lambda6584}$ $\ion{He}{II}$\,$\mathrm{\lambda4686\AA}$,  H$\alpha$, and H$\beta$)  that can separate emission-line regions excited by stellar photoionizing continuum, power-law photoionizing continuum, or shocks.   
  Therefore, these diagrams, known as Baldwin-Phillips-Terlevich (BPT) diagrams, were able to discriminate between star-forming galaxies (SFGs) and galaxies dominated by AGN activity. At the same time, a third class of galaxies was recognized by \citet{heckman1980optical} on the basis of their relatively stronger lower-ionisation lines  (Low-Ionisation Nuclear Emission line Regions; LINERs). The format of the BPT diagrams that are typically used today was refined by  \citet{1987ApJS...63..295V} to involve the $\log$([\ion{O}{III}]$\mathrm{\lambda5007\AA}/$H$\beta$) emission-line intensity ratios plotted against one of the $\log($[\ion{N}{II}]$\mathrm{\lambda6584}/$H$\alpha$), $\log$([\ion{S}{II}]$\mathrm{\lambda\lambda6716,6731}/$H$\alpha$),  $\log$([\ion{O}{I}]$\mathrm{\lambda6300}/$H$\alpha$) emission-line intensity ratios, and they can discriminate between all three classes of objects (SFGs, LINERs, AGN).

{\AZ However, the  exact demarcation between SFGs and AGNs is  generally defined empirically and hence it is subject to considerable uncertainty.   Based on stellar population synthesis and photoinization models \citet{kewley2001theoretical} introduced a maximum 'starburst' line on the BPT diagrams which defines the upper bound for the SFGs}. Driven by the fact that AGN and SFGs observed in the  Sloan Digital Sky Survey (SDSS; \citealt{2000AJ....120.1579Y}) show two distinct loci extending below the demarcation line of \citet{kewley2001theoretical}, a new empirical upper bound for the SFGs  was put forward by \citet{kauffmann2003host} in order to distinguish the pure SFGs.  The objects between this new empirical SFG line and the demarcation line of \citet{kewley2001theoretical} belong to the class of Composite galaxies (also referred to as Transition objects in previous studies; e.g. \citealt{ho1997search}). {\AZ The spectra of these Composite galaxies have been traditionally interpreted as the result of significant contributions from both AGN and star-forming activity, although, more recently it has been proposed that their strong high-excitation lines could be the result of shocks \citep[e.g.][]{2014Rich}.}
 {\new{Based on the density of the objects in the 2-dimensional diagnostic diagrams,}} \citet{kewley2006host} introduced another empirical line for distinguishing  Seyferts and LINERs. More recently, \citet{shi2015support} {\AZ explored other emission-line} intensity ratios {\AZ{that}} could improve the classification. They used support vector machines to test the classification accuracy using a dataset  of galaxies  classified as either SFG, AGN, or Composite based on \citet{kauffmann2003host}.

The currently used classification scheme suffers from {\AZ a significant drawback}. The use of multiple diagnostic diagrams  independently of one another often gives contradicting classifications for the same galaxies  \citep[e.g.][]{ho1997search}. According to \citet{kewley2006host}, $8\%$ of the galaxies in their sample are characterised as ambiguous in that they were  classified as {\AZ{belonging to}} different classes {\AZ{based on}} at least two diagnostic {\AZ{diagrams.}} {\VS For clarity, {\new{throughout this paper,}} we use the term contradicting to emphasise that the different 2-dimensional diagnostics can give different classifications.}  {\AZ{Such {\VST contradictions} arise because}} BPT diagrams are projections of a complex multi-dimensional space onto 2-dimensional planes. This limits the power of {\AZ this diagnostic tool and may lead to inconsistencies between the different diagnostic diagrams}. 
Moreover, the number of extragalactic emission-line objects for which accurate spectra are available has grown rapidly in recent years, especially with the advent of the SDSS.  This massive dataset reveals inconsistencies between the theoretical and empirical upper bounds and the actual distribution of the observed line ratios for the different classes \citep[e.g.][]{kauffmann2003host}. 
 
This limitation of the existing approach gives rise to the question of whether we can use a multidimensional data-driven method to effectively classify the galaxies. Recently, \citet{vogt2014galaxy}, generalised the diagnostics originally proposed by \citet{kewley2006host} by providing multi-dimensional surfaces {\new{in different groups of diagnostic lines}} that separate different activity classes. These, however, do not include the standard BPT diagnostic ratios. {\newtwo{Similarly, \citet{deSouza17} explore the use of Gaussian mixture models for the activity classification of galaxies in the 3-dimensional parameter space defined by the [\ion{O}{III}]$/$H$\beta$,  [\ion{N}{II}]$/$H$\alpha$, line ratios and the H$\alpha$ equivalent width (EW(H$\alpha$)). }}
 
  In this article we propose a classification scheme, the soft data-driven allocation  (SoDDA) method,  which is based on the clustering of galaxy emission-line ratios in the 4-dimensional space defined by  the [\ion{O}{III}]$/$H$\beta$,  [\ion{N}{II}]$/$H$\alpha$, [\ion{S}{II}]$/$H$\alpha$, and [\ion{O}{I}]$/$H$\alpha$ ratios. This is motivated by the clustering of the SFG, AGN, and LINER loci on the 2D projections of the emission-line diagnostic diagrams. Our classification scheme arises from a model that specifies the joint distribution of the emission-line ratios of each galaxy class to be a finite mixture of multivariate Gaussian (MG) distributions. Given the emission line ratios of each galaxy, we  compute its posterior probability to belong to each galaxy class. This allows us to achieve a soft clustering {\new{without hard separating boundaries between the different classes.}} A similar approach was successfully implemented by \citet{mukherjee1998three} in another clustering problem in which they used a mixture of MG distributions {\VS to discriminate between distinct classes of gamma-ray bursts.}
 
 This paper is organised as follows. In Section 2 we describe the proposed methodology. Section 3 discusses the  implementation of the method on galaxy spectra from the SDSS DR8, and Section 4 compares our multidimensional data driven classification scheme with the {\new{commonly used diagnostic }} proposed by \citet{kewley2006host}. Section 5 introduces multidimensional linear decision boundaries that we compare in terms of their prediction accuracy with both the SoDDA and the scheme of \citet{kewley2006host}. In Section 6 we review our results and discuss further research directions.

\section{Clustering Analysis}

 Cluster analysis is a statistical method that aims to partition a dataset into subgroups so that the members within each subgroup are more homogeneous (according to some criterion) than the population as a whole. In this article we employ a class of probabilistic (model-based) algorithms that assumes that the data are an identically and independently distributed (i.i.d.) sample from a population described by a density function, which is taken to be a mixture of component density functions. Finite mixture models have been studied extensively as a clustering technique  \citep{wolfe1970pattern}. It is common to assume that the mixture components are all from the same parametric family, such as the Gaussian. The use of mixture models arises naturally in our problem, since the population of galaxies is made up of several homogeneous {\new{and often overlapping}} subgroups {\new{from a spectroscopic perspective}}: SFGs, Seyferts, LINERs and Composites.

 \citet{fraley2002model} proposed a general framework to model a population as a mixture of $K$ subpopulations. Specifically, let $x_i$ be a vector of length $p$ containing measurements of object $i$ ($i=1,...,n)$ from a population. In our application the $x_i$ tabulates the $p=4$ emission line ratios for galaxy $i$. A finite mixture model expresses the likelihood of $x_i$ as:
 
\begin{equation}
p(x_i|\theta ,\pi) =\sum _{k=1}^K \pi _k f_k(x_i|\theta_k),  
\end{equation}

\noindent where $f_k$ and $\theta _k$ are the probability density and parameters for the distribution of  subpopulation $k$, and $\pi _k$ is the relative size of  subpopulation $k$, with $\pi _k \geq 0$ and $\sum_{i=1}^K \pi _k =1$.  Given a sample of n independent galaxies $x=(x_1,x_2,...,x_n)$, the joint density can be expressed as:
 
\begin{equation}
p(x|\theta ,\pi) =\prod_{i=1}^n\sum _{k=1}^K \pi _k f_k(x_i|\theta_k),  
\end{equation}

\noindent where $\theta=(\theta_1,...,\theta_K)$ and $\pi =(\pi_1 ,...,\pi_K)$.

\subsection{Estimating the parameters of a finite mixture model}
 \citet*{dempster1977maximum} propose a framework that can be used to  compute the maximum likelihood estimators (MLE) in finite mixture models using the Expectation-Maximization (EM) algorithm. {\VST We {\new{denote }} } the unknown parameters as $\phi=(\theta,\pi)$. The MLE is $\phi ^\star=\text{argmax} _ \phi p(x \mid \phi)$, where  argmax$_\phi$ is an operator that extracts the value of $\phi$ that maximises the likelihood function, $p(x \mid \phi)$. The EM algorithm   is an iterative method for computing the MLE.  
 
 In the context of finite mixture models, \citet{dempster1977maximum}  introduced an unobserved   vector  $z$ ($n\times K$), where $z_{i \bullet}$ is the indicator vector of length $K$ with $z_{ik}=1$ if object $i$ belongs to  subpopulation $k$ and $0$ otherwise. Because the $z_{i \bullet}$ are not observable, they are called latent variables. In this case they specify to  which subpopulation each galaxy belongs.  Given a statistical model consisting  of observed data $x$, a set of unobserved latent data $z$, and a vector of unknown parameters $\phi=(\theta,\pi)$, the EM algorithm iteratively performs alternating   expectation (E) and maximisation (M) steps: \\
 \noindent E-step: Compute $Q(\phi |\phi ^{(t)})=\text{E}[\log\: p( x,z|\phi)|x,\phi ^{(t)}]$, \\

\noindent M-step: Set $\phi ^{(t+1)} =\text{argmax} _{\phi } \:Q(\phi |\phi ^{(t)}) $, \\

\noindent where the superscript $t$ indexes the iteration, and E[.]  is the weighted mean evaluated by marginalising over all possible values of $z$. The EM algorithm enjoys stable convergence properties, in that the likelihood, $p(x|\phi)$, increases in each iteration and  the algorithm is known to converge to a stationary point of $p(x|\phi) $, which is typically a local maximum.

The joint distribution $p(x,z| \theta , \pi)$ can be factorised as $p(x,z| \theta , \pi)=p(z| \theta , \pi)p(x|z, \theta , \pi)$, where $p(z| \theta , \pi)$ is a product of $n$ multinomial distribution $p(z| \theta , \pi)=\prod_{i=1}^n\prod _{k=1}^K \pi _k ^{z_{ik}} $. Conditional on $z_{ik}=1$, $p(x_i) = f_k(x_i|\theta _k)$. The logarithm of the conditional distribution of $x$ and $z$ given $(\theta,\pi)$,  i.e. the log-likelihood,   is:

\begin{equation}
\ell (\theta, \pi  |x,z) =\log p (x,z \mid \theta, \pi  )=\sum_{i=1}^n \sum _{k=1}^K z_{ik} \log [\pi _k f_k(x_i|\theta _k) ].
\label{eq:likelihood}
\end{equation}

The E-step requires us to compute the conditional expectation of Equation~\ref{eq:likelihood} given $(\theta^{(t)},\pi^{(t)})$. Because Equation~\ref{eq:likelihood} is linear in the components of each $z_{i \bullet}$, it suffices to compute the conditional expectation of the components of each $z_{i \bullet}$ given  $x$ and  $(\theta ^{(t)}$, $\pi ^{(t)})$. This is  the conditional probabilities of $i$ belonging to subpopulation $k$ given $(\theta ^{(t)}$, $\pi ^{(t)})$. More specifically: 

\begin{equation}
\text {E}[z_{ik}|\theta ^{(t)},\pi ^{(t)},x] = \frac{ \pi_k ^{(t)} f_k(x_i|\theta _k ^{(t)})} {\sum_{k=1}^K \pi_k ^{(t)} f_k(x_i|\theta _k ^{(t)})} =\gamma(z_{ik})
\label{Eq:Estep}
\end{equation}

The M-step requires us  to maximise the conditional expectation of Equation~\ref{eq:likelihood} with respect to $\pi $ and $\theta $, i.e. to maximise $\sum_{i=1}^n \sum_{k=1} ^K \gamma(z_{ik}) \log  [\pi _k f_k(x_i|\theta _k) ]$. The particular form of the M-step depends on the choice of density distributions, $f_k$, for the subpopulations. Here we assume MG distributions for each subpopulation.

MG mixture models can be used for data with varying structures due to the flexibility in the definition of  variance matrices. The density of the MG distribution for subpopulation $k$ is: 

\begin{equation}
f_x(x_i) = \frac{1}{\sqrt{(2\pi )^p | \Sigma_k | }} \exp \Big ( -\frac{1}{2} (x_i-\mu_k )^{T } \Sigma_k ^{-1} (x_i-\mu_k ) \Big ).
\label{eq:MG}
\end{equation}

 The EM formulation for an MG mixture is presented in detail in \citet{dempster1977maximum}. The E-step has the same formulation as in Equation~\ref{Eq:Estep}, with $f_k$ given in Equation~\ref{eq:MG} with 
 $\theta _k=(\mu_k, \Sigma _k)$, where $\mu_k$ represent the means and $\Sigma_k$ the covariance matrices of the $x_i$ line ratios for galaxies in subpopulation $k$. For the M-step, the updates of the parameters have closed form solutions \citep{bilmes1998gentle}, 

\begin{align}
\pi_k^{(t+1)}&=\frac{1}{n}\sum_{i=1}^n  \gamma(z_{ik})\\
\mu_k^{(t+1)}&=\frac{\sum_{i=1}^n x_i  \gamma(z_{ik})}{\sum_{i=1}^n \gamma(z_{ik})}\\
\Sigma_k^{(t+1)}&=\frac{\sum_{i=1}^n  \gamma(z_{ik}) (x_i-\mu_k^{(t+1)})(x_i-\mu_k^{(t+1)})^T}{\sum_{i=1}^n \gamma(z_{ik})}.
\end{align}

\noindent   We implement this EM algorithm using the {\tt scikit-learn} Python library\footnote{http://scikit-learn.org/stable/}  under the constraint that the covariance matrices are full rank, and the diagonal elements cannot be smaller than $10^{-3}$ to avoid overestimation, i.e. converging to a small number of data points.  Because this algorithm can be sensitive to the choice of starting values, we routinely rerun it with 5 different randomly selected sets of starting values. {\new{The values of the likelihood for the different starting values differ less than $0.5\%$}}.  We choose the value among the 5 converged points with the largest likelihood to be the MLE, denoted $(\pi^\star, \mu^\star, \Sigma^\star)$.

\subsection{Choosing the value of $K$}
\citet{fraley2002model} point out that mixtures of MG distributions are appropriate if the subpopulations are   centred at the means, $\mu _k$, with increased density for data closer to the means. As a result, the practical use of MG mixture models could be limited if the  data  exhibit non-Gaussian features, including asymmetry, multi-modality and/or heavy tails. In the SDSS {\AZ DR8} dataset that we examine, it is apparent that the subpopulations exhibit non Gaussians characteristics such as convexity,  skewness and  multimodality. In order to account for these non-Gaussian features, we use a mixture of MG distributions with $K$ considerably larger than the actual number of galaxy classes. In this way, we represent each galaxy class by a mixture of several MG subpopulations. This allows a great deal of flexibility in the class-specific distributions of emission line ratios. With the fitted (large $K$) MG mixture in hand we can then perform hyper-clustering of the $K$ MG subpopulations so as to concatenate them into clusters representing the four desired galaxy classes.

The number $(K>>4)$ of MG subpopulations that we  fit to our data is chosen using the Bayesian Information Criterion (BIC) of \citet{schwarz1978estimating} and the gap statistic \citep{tibshirani2001estimating}.  BIC is a model selection criterion based on the maximum log-likelihood obtained with each possible value of $K$, and penalised by the increased complexity associated with more subpopulations . More specifically, it is defined as $\text{BIC}(K)=-2\cdot L^\star(K) +K\log(n)$, where $L^\star(K) =p(x \mid \theta ^\star (K), \pi^ \star (K) )$ is the maximised value of the likelihood when the number of subpopulations is fixed at $K$. The value of $K$ with the lowest BIC is preferred.  The gap statistic compares the normalised intra-cluster distances between points in a given cluster, $W_K$, for different total number of subpopulations $K$, with a null reference distribution obtained assuming data with no obvious clustering. The null reference distribution is generated by sampling uniformly from the original datasets bounding box multiple times. The estimate for the optimal number of subpopulations $K$ is the value for which the $W_K$  falls the farthest below the reference curve. 

SoDDA accomplishes the hyper-clustering of the $K$ subpopulations  into the four galaxy classes using  the classification scheme of \citet{kewley2006host}. More specifically, we treat the fitted subpopulations means $(\mu_1^\star,...,\mu_K^\star)$ as a dataset  and  classify them into the four galaxy classes.  For example, suppose we fit 10 MG distributions  and the means of the distributions $1,3$ and $5$ are classified by \citet{kewley2006host} as SFGs, then the distribution of the SFGs under SoDDA would be 
\begin{equation}
f_{\text{SFG}}(x_i)=\frac{\pi_1^\star f_1(x_i|\theta_1^\star, \pi_1^\star) +\pi_3 ^\star f_3(x_i|\theta_3^\star ,\pi_3^\star) +\pi_5^\star f_5(x_i|\theta_5^\star ,\pi_5^\star)}{\pi_1^\star +\pi_3^\star +\pi_5^\star}.
\end{equation} 

Via the allocations of the means of the $K$ subpopulations into the four galaxy classes, we have defined the distribution of the emission line ratios for each galaxy class  as a finite mixture of MG distributions. Specifically, let $f_{\text{SFG}}(x), f_{\text{LINER}}(x),f_{\text{Seyfert}}(x)$, and $f_{\text{Comp}}(x)$ be the distributions under SoDDA of the emission line ratios of SFGs, LINERs,  Seyferts and  Composites galaxies respectively. Then, given the four emission line ratios $x_i$ of a galaxy $i$, the posterior probability of galaxy $i$ belonging to class $c$  is:

\begin{align}
\rho_{ic}&=\Pr(\text{galaxy $i$ is of class $c$}) \\
&=\frac{f_c(x_i)}{\sum _z f_z(x_i)},\quad \text{for $z$ in }\{\text{SFG},\text{LINER},\text{Seyfert},\text{Comp}\} .\label{eq:posterior}
\end{align}

\section{Implementation of the classification Scheme}\label{dataset}

The SDSS provides an excellent resource of {\new{ spectra of the central regions ($\sim5.5$\,kpc for $\mathrm{z<0.1}$) of galaxies covering all different activity types  \citep[e.g.][]{kauffmann2003host}}}. For the definition of our multi-dimensional activity diagnostics we use the "galspec"  database of spectral-line measurements from the Max-Plank Institute for Astronomy and Johns Hopkins University group. We used the version of the catalog made publicly available through the SDSS Data Release 8 \citep{aiharaa,aiharab,eisenstein}, which contains 1,843,200 objects. The spectral-line measurements are based on single Gaussian fits to star-light subtracted spectra, and they are corrected for foreground Galactic absorption \citep{tremonti2004,kauffmann2003host,2004MNRAS.351.1151B}. Since the same catalog has been used for the definition of the two-dimensional and multi-dimensional diagnostics of \citet{kauffmann2003host} and \citet{vogt2014galaxy} respectively, it is the best benchmark for testing the SoDDA. 
{\new{ Before proceeding with our analysis we applied the corrections on the line-measurement errors reported in \cite{juneau2014}, and we corrected the flux of the H$\beta$ line following \citet{groves2012}.
From this catalog we selected all objects satisfying the following criteria, which closely match those used in the reference studies of \citet{kauffmann2003host} and \citet{kewley2006host}:
\begin{itemize}
\item{{\sc{RELIABLE=1}} "galspec"  flag. }
\item{No warnings for the redshift measurement (\sc{Z\_WARNING=0}).}
\item{Redshift between 0.04 and 0.1.}
\item{Signal-to-noise ratio (SNR) greater than 3 on each of the strong emission-lines  used in this work

: H$\alpha$, H$\beta$, [\ion{O}{III}]$\mathrm{\lambda5007}$ [\ion{N}{II}]$\mathrm{\lambda6584}$, [\ion{S}{II}]$\mathrm{\lambda\lambda6716,6731}$. This ensures the use of reliable line flux measurements for our analysis.}
\item{The continuum near the H$\beta$ line has SNR$>3$. }
\item{Ratio of H$\alpha$ to corrected H$\beta$ greater than the theoretical value 2.86 for star-forming galaxies. This excludes objects with problematic starlight subtraction and errors on the line measurements (c.f. \citealt{kewley2006host})}
\end{itemize}
The final sample consists of 130,799 galaxies, and it provides a direct comparison with the reference diagnostics of \citet{kauffmann2003host} and \citet{kewley2006host} which have used very similar selection criteria.
Given the difficulty in correcting for intrinsic extinction in the cases of Composite and LINER galaxies we do not attempt to apply any extinction corrections (apart from the requirement for the galaxies to have positive Balmer decrement). }}

\begin{figure}
\includegraphics[width=\columnwidth]{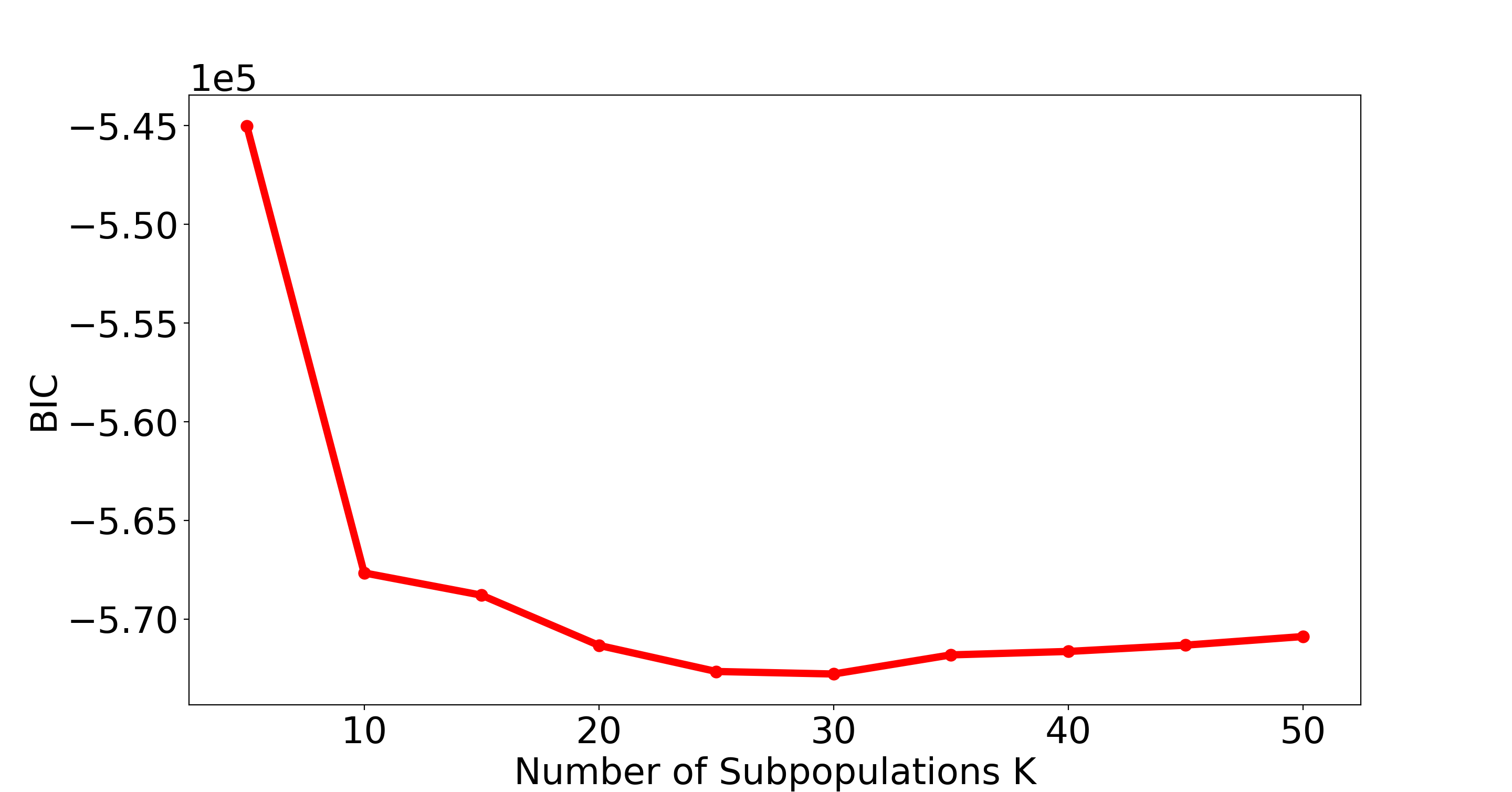}
\caption{	The Bayesian Information Criterion (BIC) computed over a grid of values of $K$ ({\AZ in} increments of 5) using the data of the SDSS DR8. The BIC is a model selection criterion based on the log-likelihood; the model with the lowest BIC {\AZ value}  is preferred, {\AZ indicating that in this case the optimal number of subpopulations is} $K=25$.} 
\label{fig:test2}
\end{figure}

\begin{figure}
\includegraphics[width=\columnwidth]{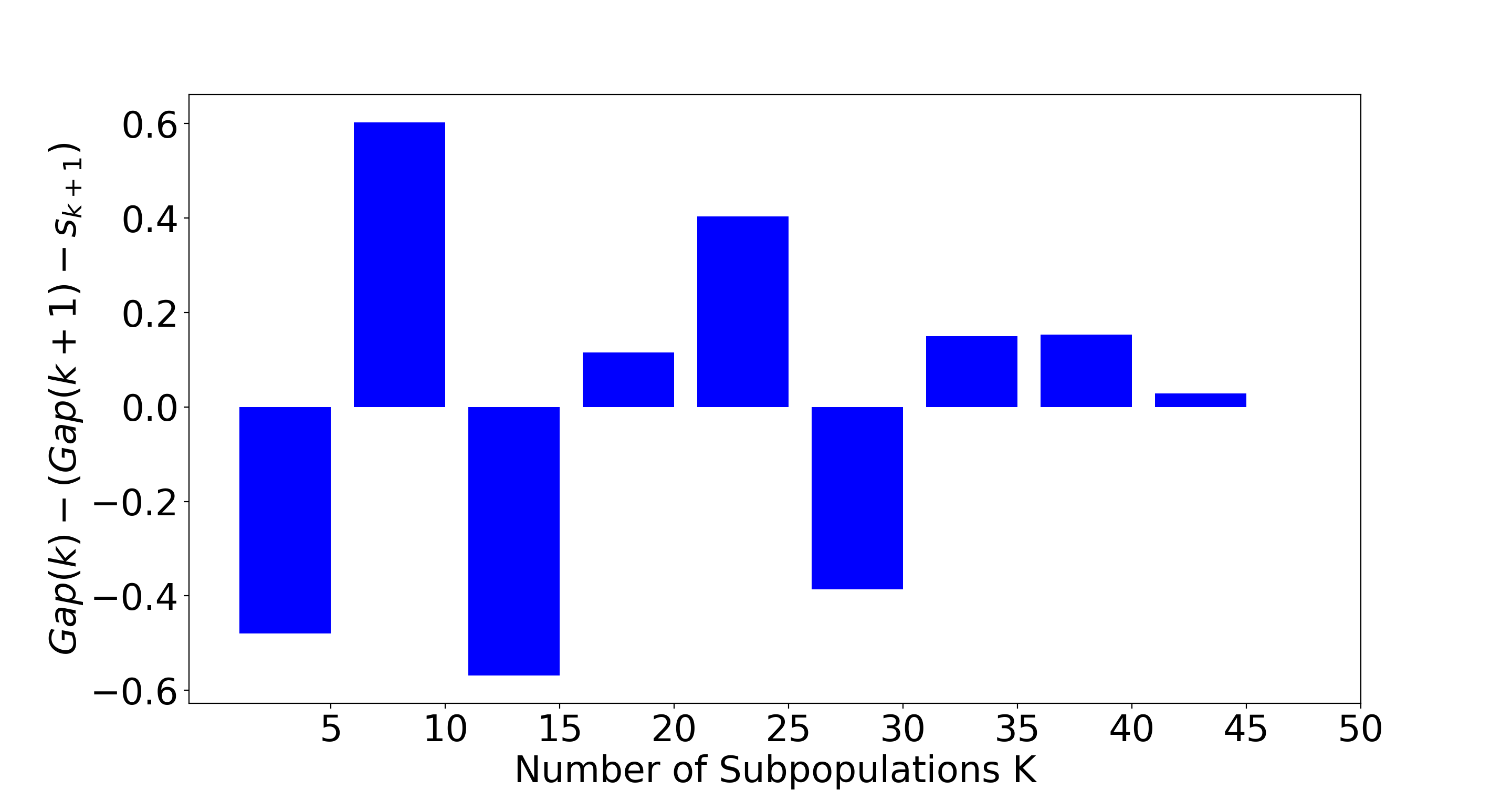}
\caption{The Gap statistic computed over a grid of values of $K$ ({\AZ in} increments of 5) using the data of the SDSS DR8. The Gap statistic compares the intra-subpopulation distances between points in a given subpopulation with a null reference distribution of the data, {\AZ i.e.,} a distribution with no obvious clustering. {\AZ This figure} shows that the smallest value of $K$ for which the data measure
exceeds the randomly generated measure is $K=10$.} 
\label{fig:test1}
\end{figure}

 We apply the BIC and gap statistic for  values of $K$ ranging from 5 to 50 in increments of 5. Figures \ref{fig:test2} and \ref{fig:test1} plot the BIC and gap statistics. BIC suggests an optimum  value of around $K=25$, while the gap statistic suggests a value of $K=10$. Since we are ultimately concatenating  the subpopulations, we err on the side of large $K$, with $K=20$, so as to capture as much detail in the data as possible without over-fitting.

\begin{figure*}

\includegraphics[width=\linewidth]{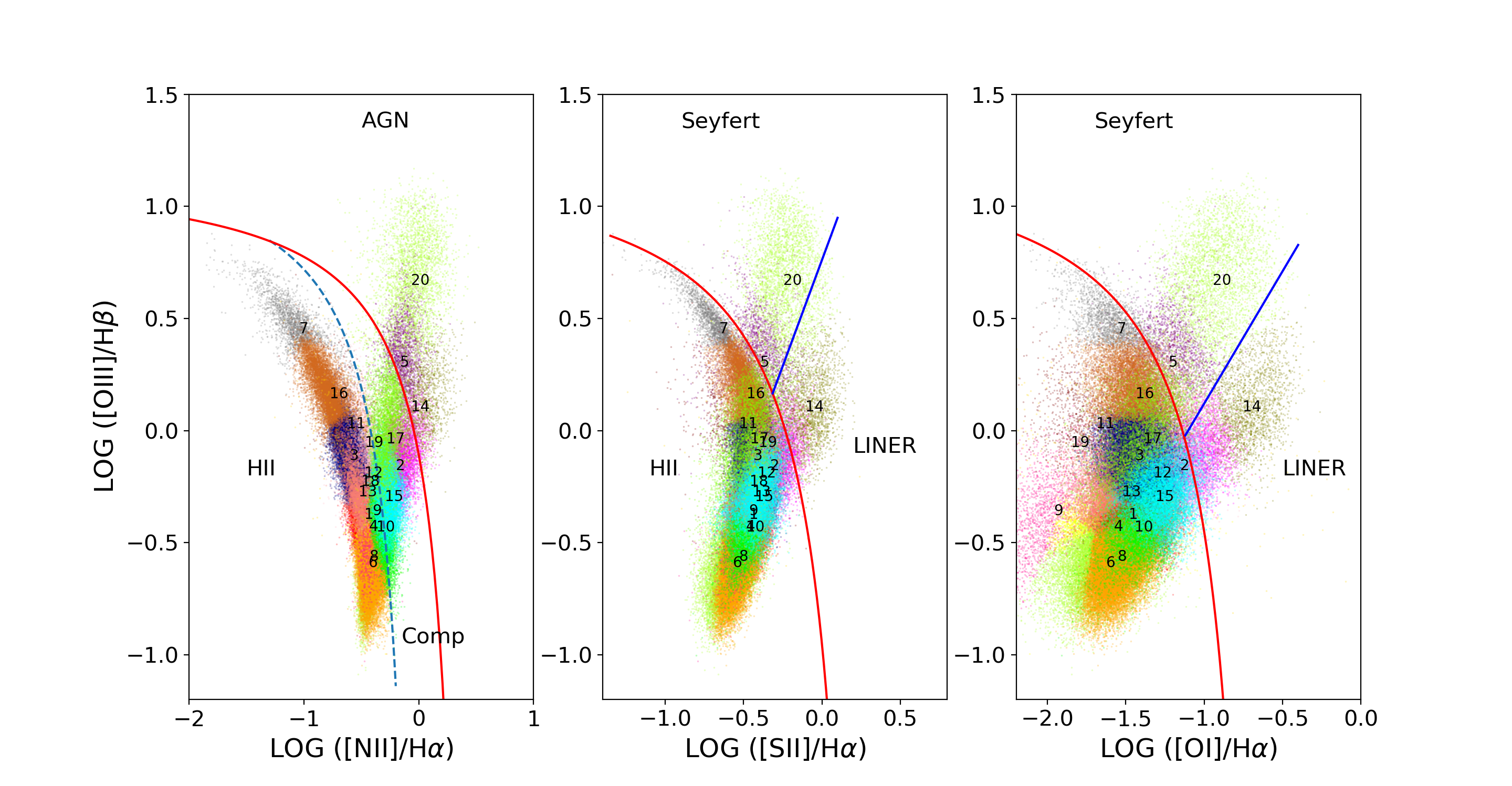}
\caption{The BPT diagnostic diagrams for the  {\VS SDSS DR8} sample; each galaxy is coloured according to its most probable allocation to one of the 20 subpopulations. The maximum 'starburst' line of \citet{kewley2001theoretical} is shown {\AZ by the} solid red line and the empirical upper bound on SFG of \citet{kauffmann2003host} is plotted as dashed blue line. The empirical line for distinguishing  Seyferts and LINERs of  \citet{kewley2006host} is depicted {\AZ by the} solid blue line.} 
\label{fig:20}
\end{figure*}

\begin{figure*}
\includegraphics[width=\linewidth]{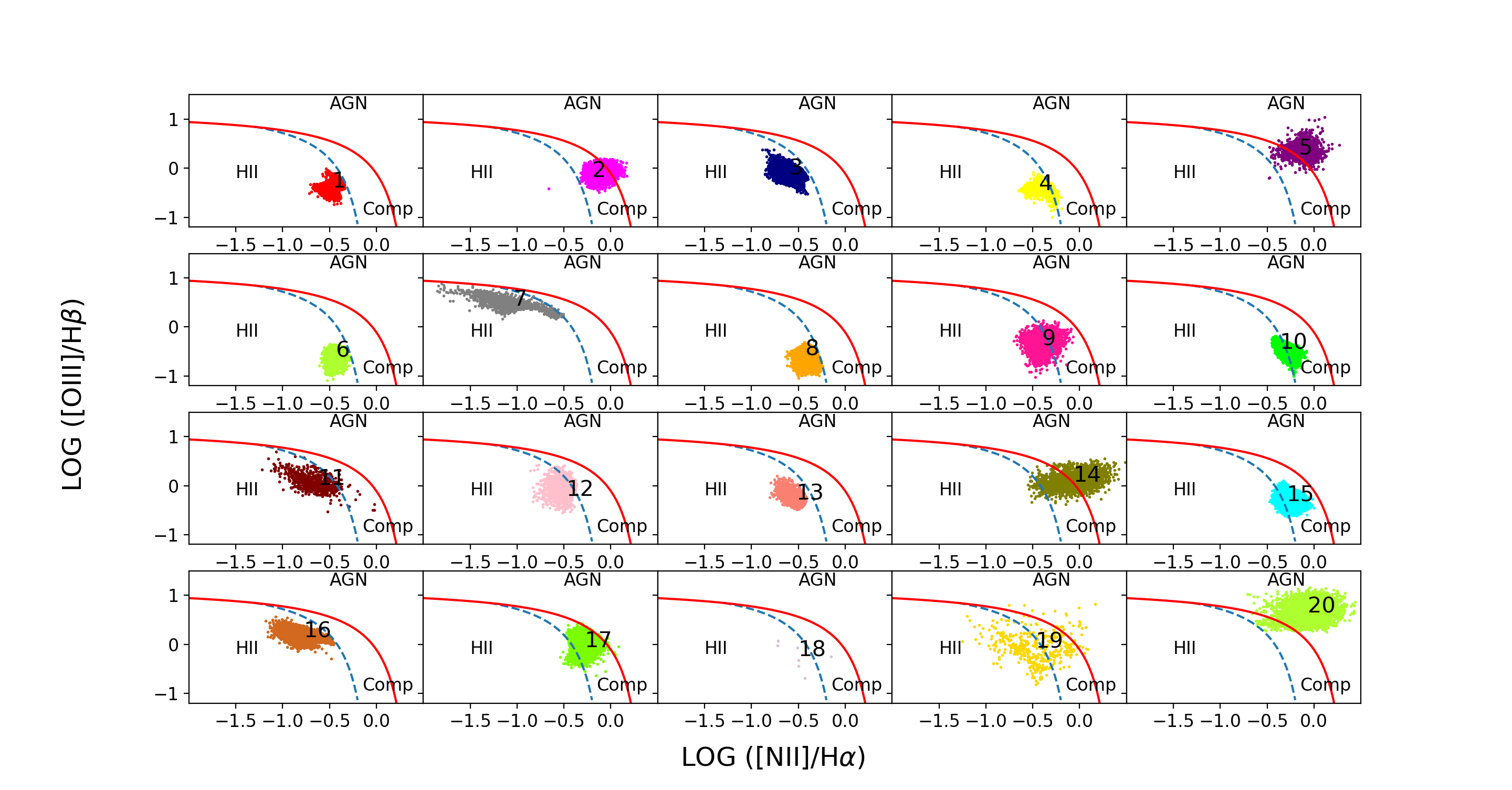}
\caption{	{\AZ The  20 subpopulations plotted on the [\ion{N}{II}]$/$H$\alpha$ vs [\ion{O}{III}]$/$H$\beta$ projection of the 4-dimensional diagnostic diagram. The subpopulations are numbered following the scheme in Figure~\ref{fig:20}. This figure shows the spatial extent of each subpopulation and their location with respect to the standard diagnostic lines in the [\ion{O}{III}]$/$H$\beta$ diagram. Since these are 2-dimensional projection of the 4-dimensional distribution in each subpopulation, they only give an indication of the extent and location of each subpopulation.
}}
\label{fig:comment7}
\end{figure*}

Figure~\ref{fig:20} displays the BPT diagnostic diagrams for {\VS SDSS DR8} with each point colour coded according to its most probable subpopulation among the $K=20$ fit. The  means of the subpopulations are plotted for $k=1,\ldots, 20$. {\VS To visualize the spacial extent of each of the 20 subpopulation, Figure~\ref{fig:comment7} plots the [\ion{N}{II}]$/$H$\alpha$ vs [\ion{O}{III}]$/$H$\beta$ diagnostic diagram for each  subpopulation \newtwo{(Subpopulation 18 contains very few objects, mostly capturing objects with large errors in the [\ion{O}{I}]$/$H$\alpha$ ratios).} We emphasize that the full 4-dimensional geometry of the subpopulations cannot be seen in the 2-dimensional projections.}


%
 
\begin{figure*}
\centering
\includegraphics[width=0.75\linewidth]{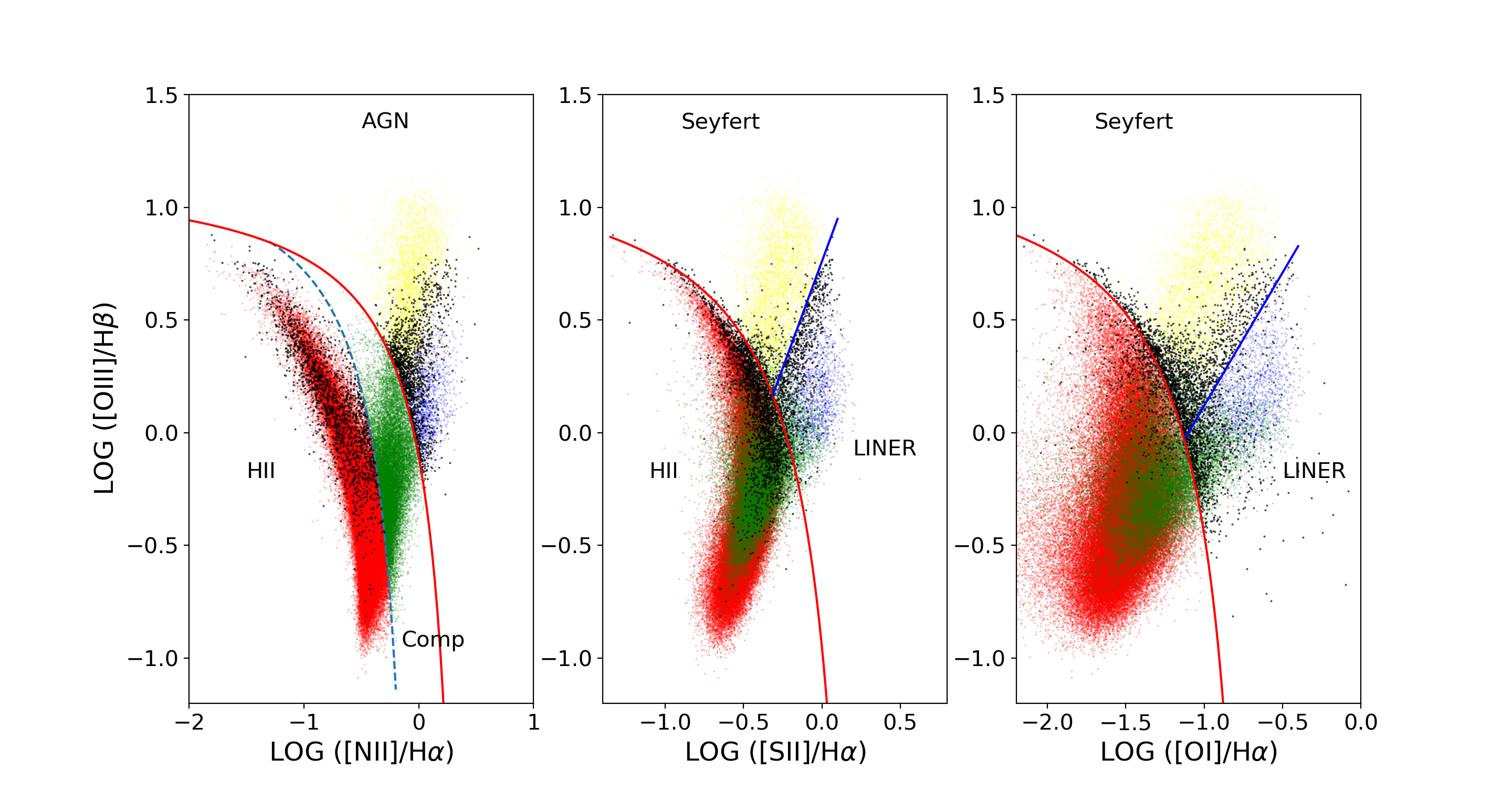}
\vspace{20pt}
\includegraphics[width=0.75\linewidth]{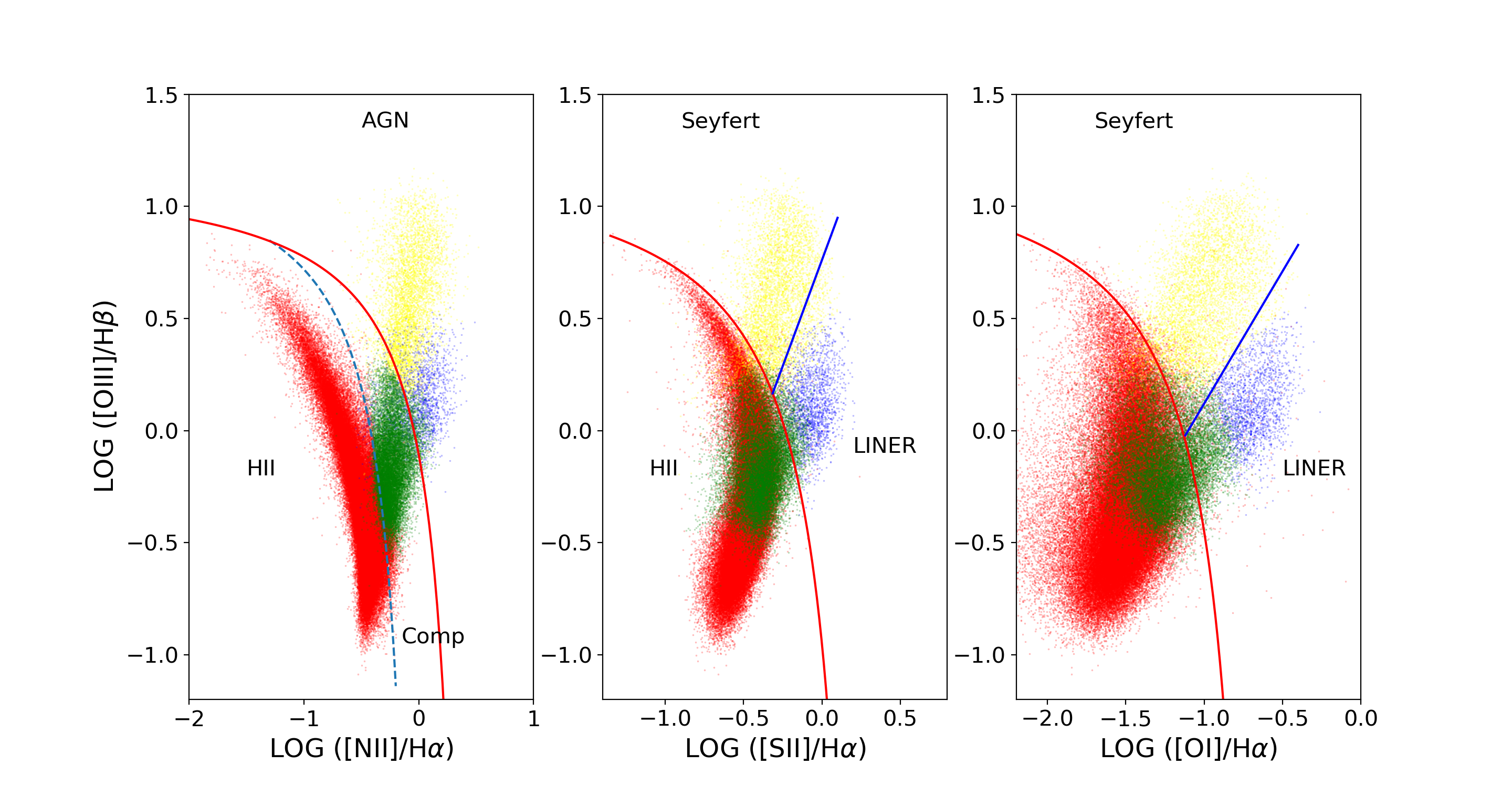}
\caption{The BPT diagrams for the galaxies in the SDSS DR8 sample, based on the  \citet{kewley2006host}  scheme (top) and SoDDA (bottom).  Each galaxy is colour coded according to its classification: red for SFGs, yellow for Seyferts, blue for LINERs, green for the Composite galaxies, and black for the Contradicting classifications. Note the lack of any contradicting classifications (black points) in the SoDDA results (bottom). For reference we also plot the  the maximum 'starburst' line of \citet{kewley2001theoretical} (solid red), the empirical upper bound on SFG of \citet{kauffmann2003host} (dashed blue), and the empirical line distinguishing  Seyferts and LINERs  (\citealt{kewley2006host}; solid blue).  3-dimensional rotating projections of the 4-dimensional diagram of the SoDDA classification (depicted in the bottom row of the figure in 2-dimensional projections) are available online: \url{http://hea-www.harvard.edu/AstroStat/etc/gifs.pdf}. The animated figures can also be found as supplementary material.}
\label{fig:class_pred}
\end{figure*}
\begin{landscape}
\begin{figure}
\begin{tabular}{ll}
\hspace{-25pt}\includegraphics[height=0.25\paperheight]{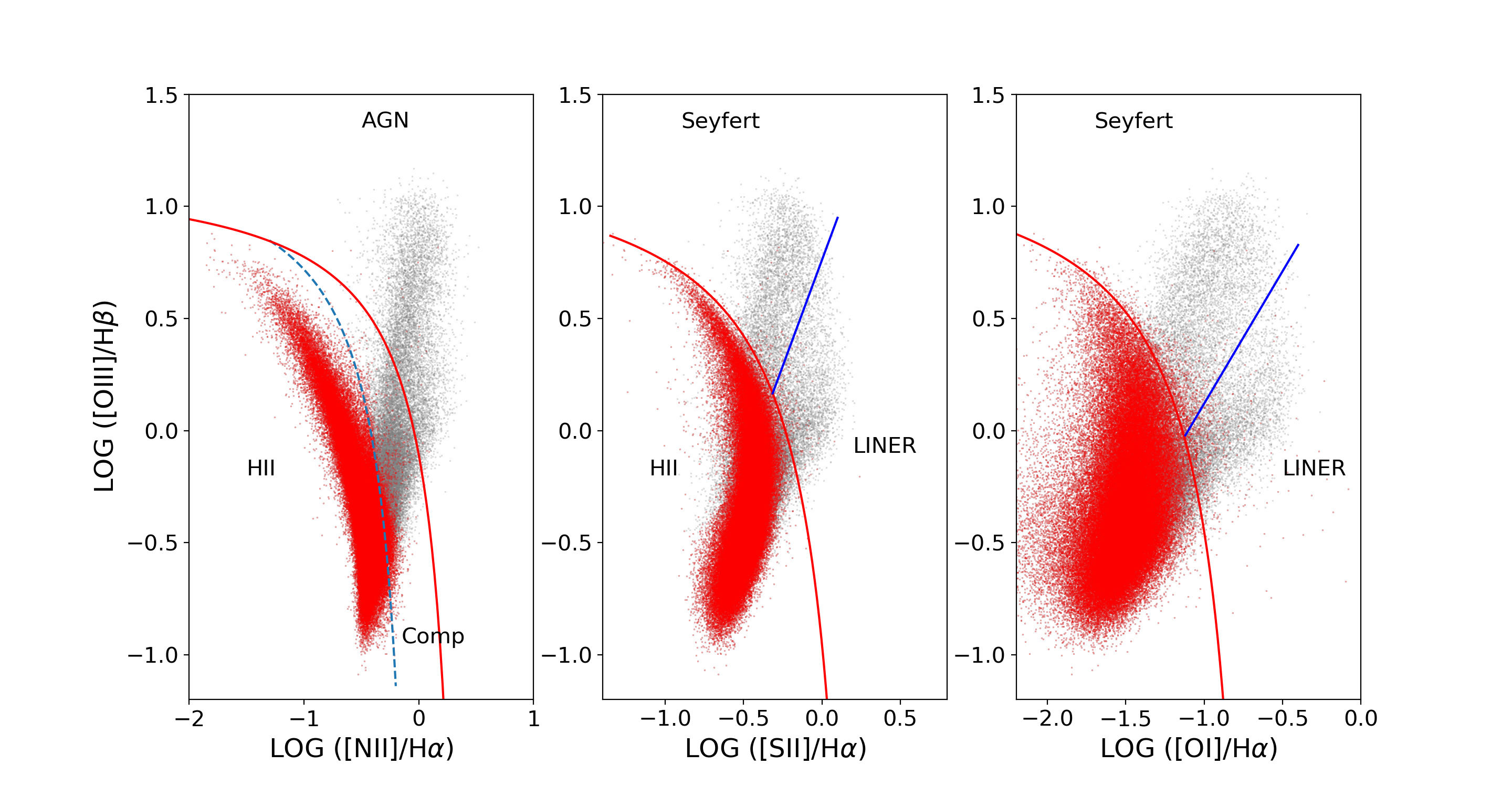}
&
\hspace{-25pt}\includegraphics[height=0.25\paperheight]{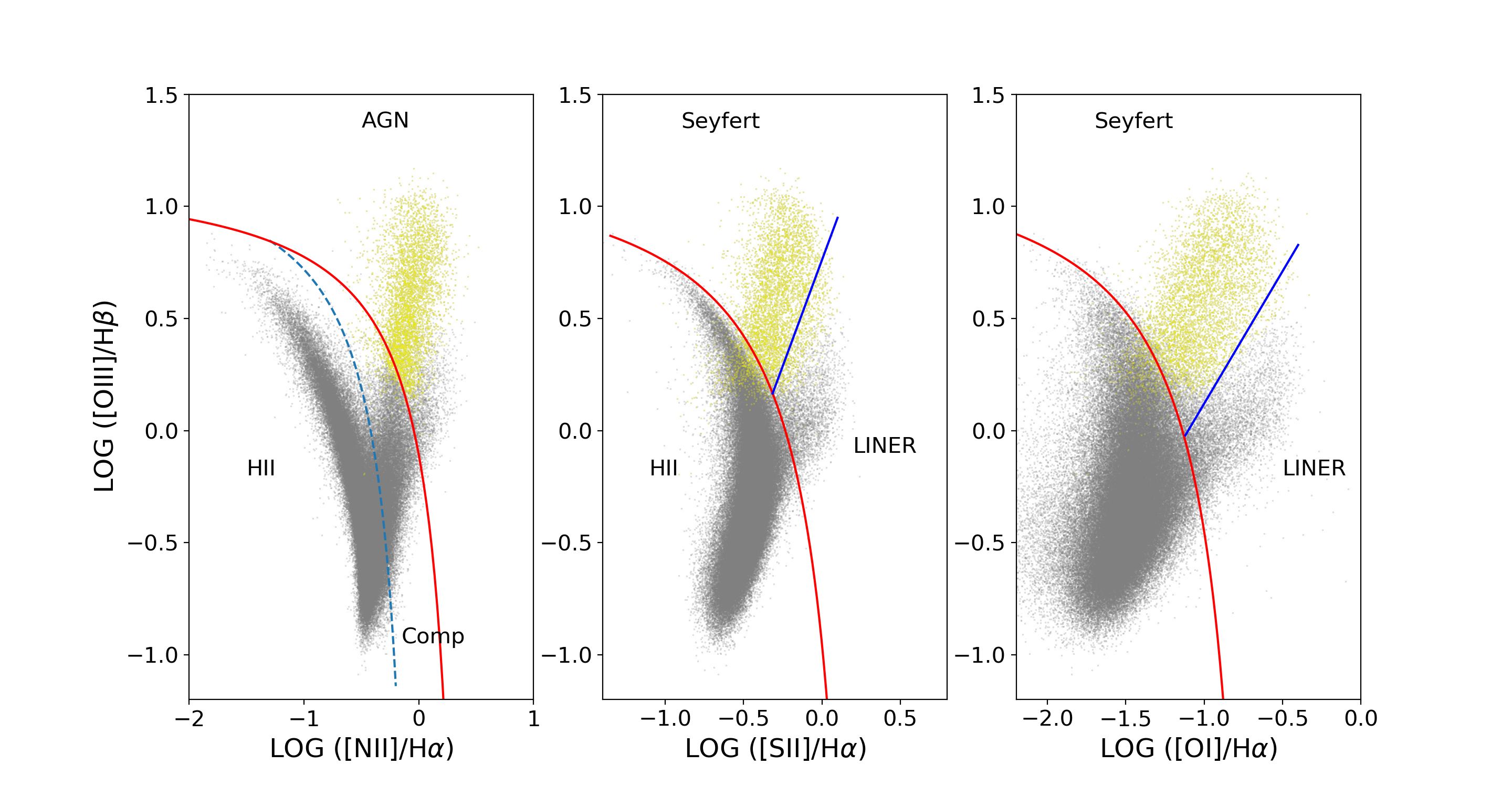}
\\
\hspace{-25pt}\includegraphics[height=0.25\paperheight]{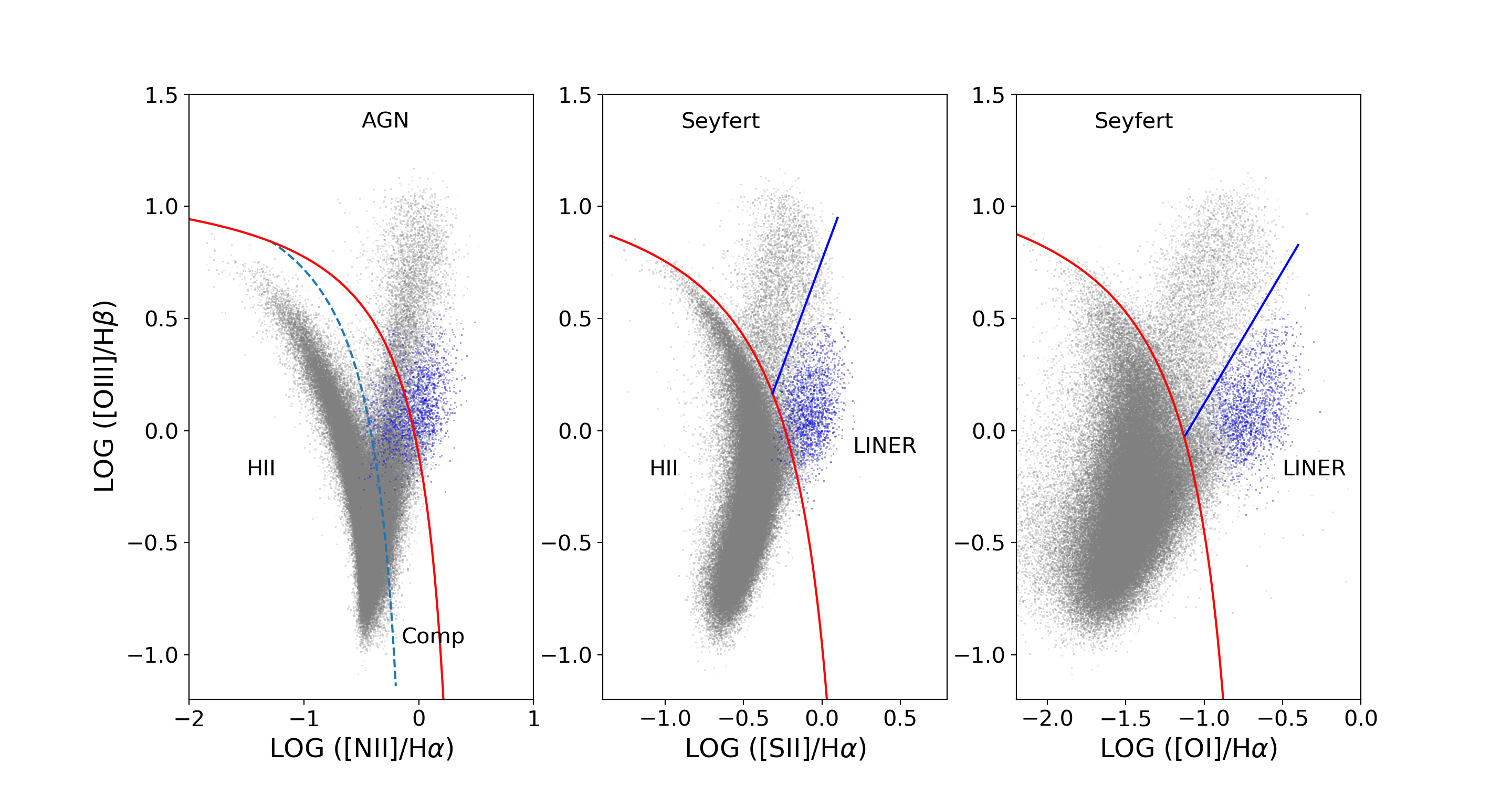}
&
\hspace{-25pt}\includegraphics[height=0.25\paperheight]{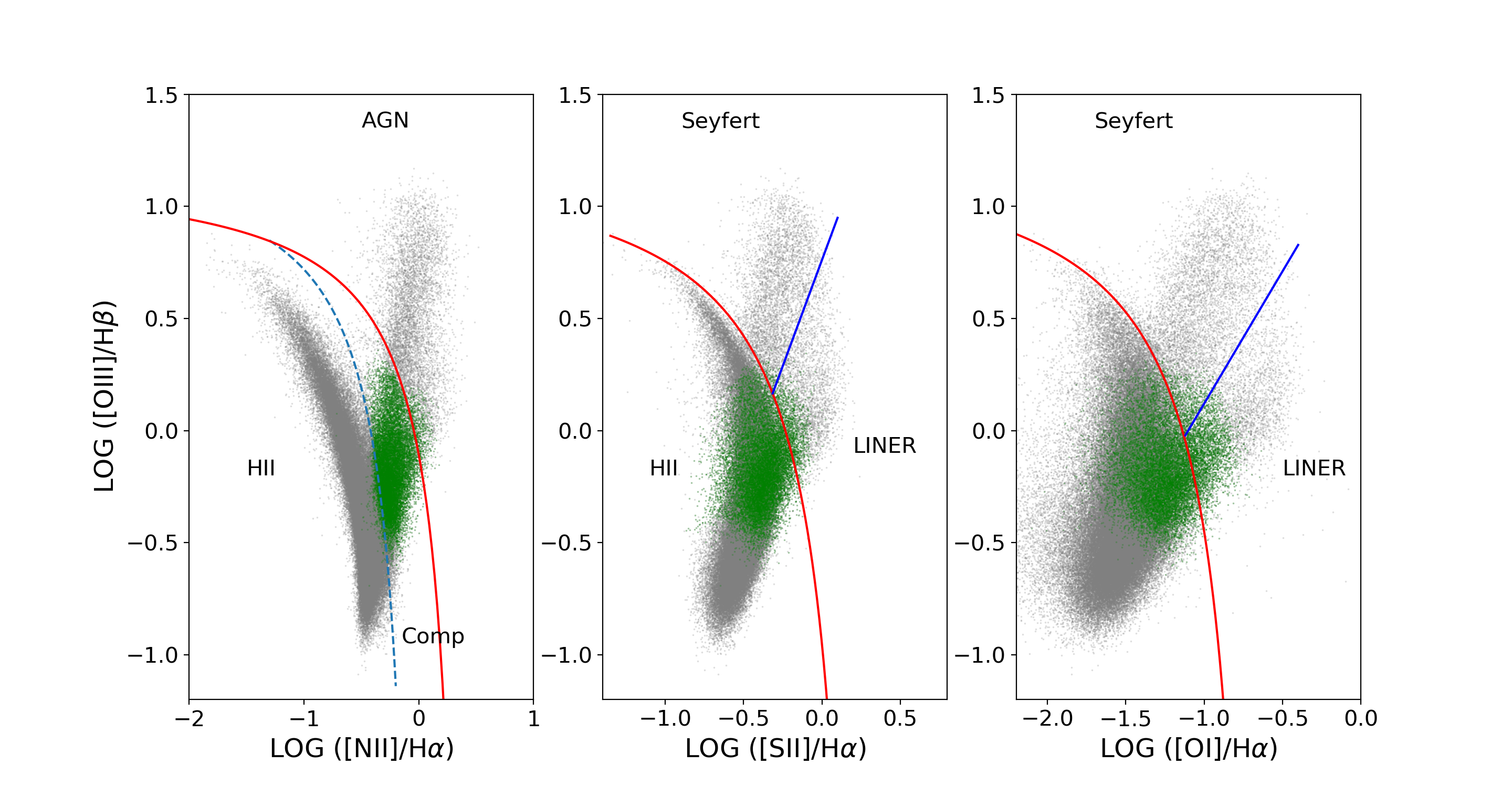}
\\
\end{tabular}
\caption{ The locus of galaxies classified into the different activity types using SoDDA plotted on the three BPT diagrams. Each set of panels shows a different class (clockwise from top left):  (a) SFGs (red), (b) Seyfert (yellow); (c) LINERs (blue), (d) Composite (green). For reference the full sample is also plotted in grey. The maximum 'starburst' line of \citet{kewley2001theoretical} is plotted as a solid red line, the empirical upper bound on SFG of \citet{kauffmann2003host} is plotted as a dashed blue line,  and the empirical line distinguishing  Seyferts and LINERs  \citep{kewley2006host} is plotted as a solid blue line.
}
\label{fig:SFG}
\end{figure}
\end{landscape}

SoDDA associates each of the {\new{20}} subpopulations with one activity class based on {\AZ the projection of their mean} on the 2-dimensional BPT diagnostic diagrams, and their location with respect to the activity-class separating lines reported in \citet{kewley2006host}. The allocations are {\AZ given} in Table~\ref{tab:initial_comp} for the 20 subpopulations means. 
{\new{All but subpopulation 5 can be clearly associated with one activity class in all three diagnostic planes.   
The mean of subpopulation 5 is located within the Seyfert class, but its extent transcends the Composite and Seyfert classes. Since the main discriminator between Composite galaxies and Seyferts is the [\ion{N}{II}]$/$H$\alpha$ diagnostic  and the mean of this subpopulation is clearly above  the maximum `starburst' line on the BPT diagrams introduced by \citet{kewley2001theoretical} as an upper bound of SFGs, we include Subpopulation~5  in the Seyfert class. }} After combining the 20 subpopulations to form the 4 galaxy classes as described in Table~\ref{tab:initial_comp}, we compute the posterior probability of each galaxy being a SFG, Seyfert, LINER, or Composite using Equation~\ref{eq:posterior}.  {\VS The second row in Figure~\ref{fig:class_pred} }shows the BPT diagnostic diagrams for SDSS {\AZ DR8}  with each galaxy colour coded according to its most  probable galaxy class (red for SFGs, yellow for Seyferts, blue for LINERs, and green for the Composites) under SoDDA. {\VS To highlight  the spatial extent of each cluster, we plot the BPT diagrams for each activity class (SFGs, Seyferts, LINERs and Composites) individually in Figure~\ref{fig:SFG}.}

\begin{table}
\caption{The suggested classification of the {\new{20}} subpopulations means.}
\begin{tabular}{c r}
	Class & Subpopulation ID \\
	\hline
             SFG & {\new{1,  3,  4,  6,  7,  8,  9, 10, 11, 12, 13, 16, 19 }}\\
               Seyferts & {\new{5, 20}}\\
                LINER & {\new{14}}\\
                Composites &{\new{2, 15, 17, 18}} \\
\end{tabular}
\label{tab:initial_comp}
\end{table}

\begin{figure*}
\includegraphics[width=0.9\linewidth]{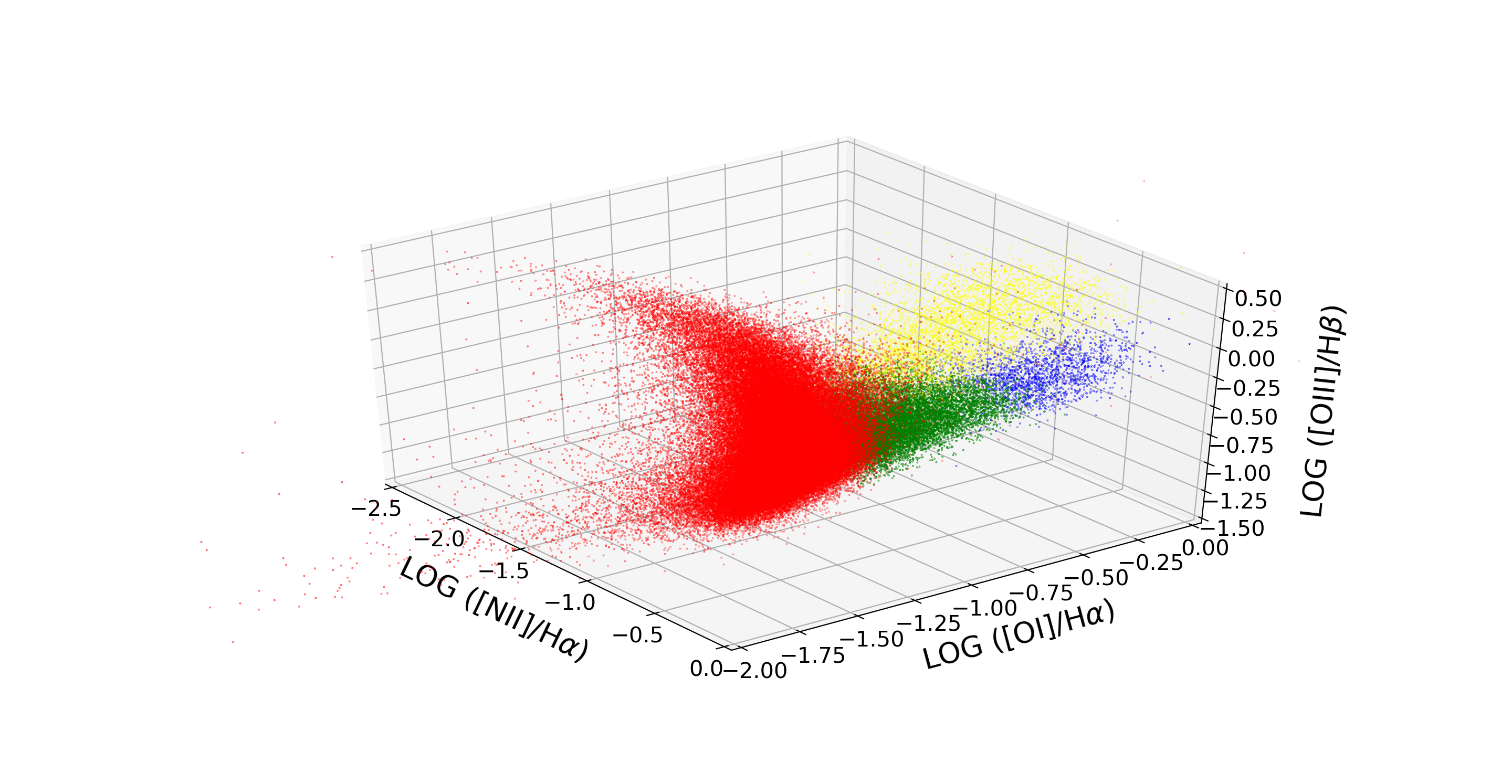}
\caption{ A 3-dimensional projection of the SDSS {\AZ DR8} sample {\AZ used in our study} on the ([\ion{N}{II}]$/$H$\alpha$, [\ion{S}{II}]$/$H$\alpha$, [\ion{O}{III}]$/$H$\beta$) volume, in which each galaxy is colour coded according to its SoDDA classification (red for SFGs, yellow for Seyferts, blue for LINERs and green for the Composites). The 3-dimensional projections illustrates the complex structure of the 4 galaxy activity classes. Each of the four 3-dimensional rotating projection of the full 4-dimensional diagram are available online: \url{http://hea-www.harvard.edu/AstroStat/etc/gifs.pdf}.}
\label{fig:class_3d}
\end{figure*}

Figure~\ref{fig:class_3d} depicts a 3-dimensional projection of the SDSS {\AZ DR8} sample on the ([\ion{N}{II}]$/$H$\alpha$, [\ion{S}{II}]$/$H$\alpha$, [\ion{O}{III}]$/$H$\beta$) volume. This 3-dimensional projections illustrate the complex structure of the 4 galaxy activity classes. 3-dimensional rotating projections can be found at \url{http://hea-www.harvard.edu/AstroStat/etc/gifs.pdf}

{\newtwo{The data used for Figs \ref{fig:class_3d}, \ref{fig:class_pred}, \ref{fig:SFG} are presented in Table~\ref{tab:galaxy_class}. This table gives the SoDDA-based probability that each galaxy in the sample considered here belongs to each one of the activity classes, along with the galaxy's SPECOBJID, the key diagnostic line-ratios, and the activity classification based on the class with the highest probability.  Table~\ref{tab:galaxy_class} contains the details for five galaxies of the sample we used. We include the table for the entire sample in the online version of the paper. 
}}

\begin{landscape}
\begin{table}
\caption{ {\newtwo{Activity classification of the emission-line galaxies in the SDSS-DR8  based on the SoDDA. Column (1): SPECOBJID in SDSS DR8; Columns (2), (3), (4), (5): Logarithm of the diagnostic line-ratios (see \SS3); Columns (6), (7), (8), (9): Probability that a galaxy belongs to each one of the 4 activity classes based on the SoDDA analysis; Column (10): Highest-ranking activity class: 0 for SFGs, 1 for Seyferts, 2 for LINERs, and 3 for Composites. We include the table for the entire sample in the online version of the paper.}}}
\begin{tabular}{lccccccccc}
	 & \multicolumn{4}{c}{Line Ratio}  &  \multicolumn{4}{c}{SoDDA Probability} &  \\
	  SPECOBJID & $\log$([\ion{N}{II}]$/$H$\alpha$)& $\log$([\ion{S}{II}]$/$H$\alpha$) & $\log$([\ion{O}{I}]$/H\alpha$) & $\log$([\ion{O}{III}]$/$H$\beta$) &
	          SFG  &  Seyfert & LINER & Composite &  Activity Class \\
	\hline
  299491051364706304 & -0.525441 & -0.556073  & -1.623533 &  -0.621178 & 0.992937 & 0.000052  & 3.217684e-09 & 0.007011 &             0 \\   
  299492700632147968 & -0.442478    &     -0.479489  & -1.467312        & -0.572390 & 0.983635 &   0.000046 & 8.869151e-08 & 0.016319 &             0 \\ 
  299493525265868800 & -0.516100   &      -0.482621  & -1.482500      &   -0.262816 & 0.989069      & 0.000207  &7.396101e-07 & 0.010723 &             0 \\ 
  299493800143775744 &-0.665688    &     -0.392920       &   -1.630935      &   -0.081032 &  0.999946    &   0.000007 & 1.841213e-09 & 0.000048 &             0 \\ 
  299494075021682688 & -0.305985     &    -0.285281        &   -1.293723     &    -0.274226  &0.189374    &   0.006725 & 7.278570e-04 & 0.803174 &             3 \\ 
\end{tabular}
\label{tab:galaxy_class}

\end{table}
\end{landscape}

\begin{figure*}
\includegraphics[width=\linewidth]{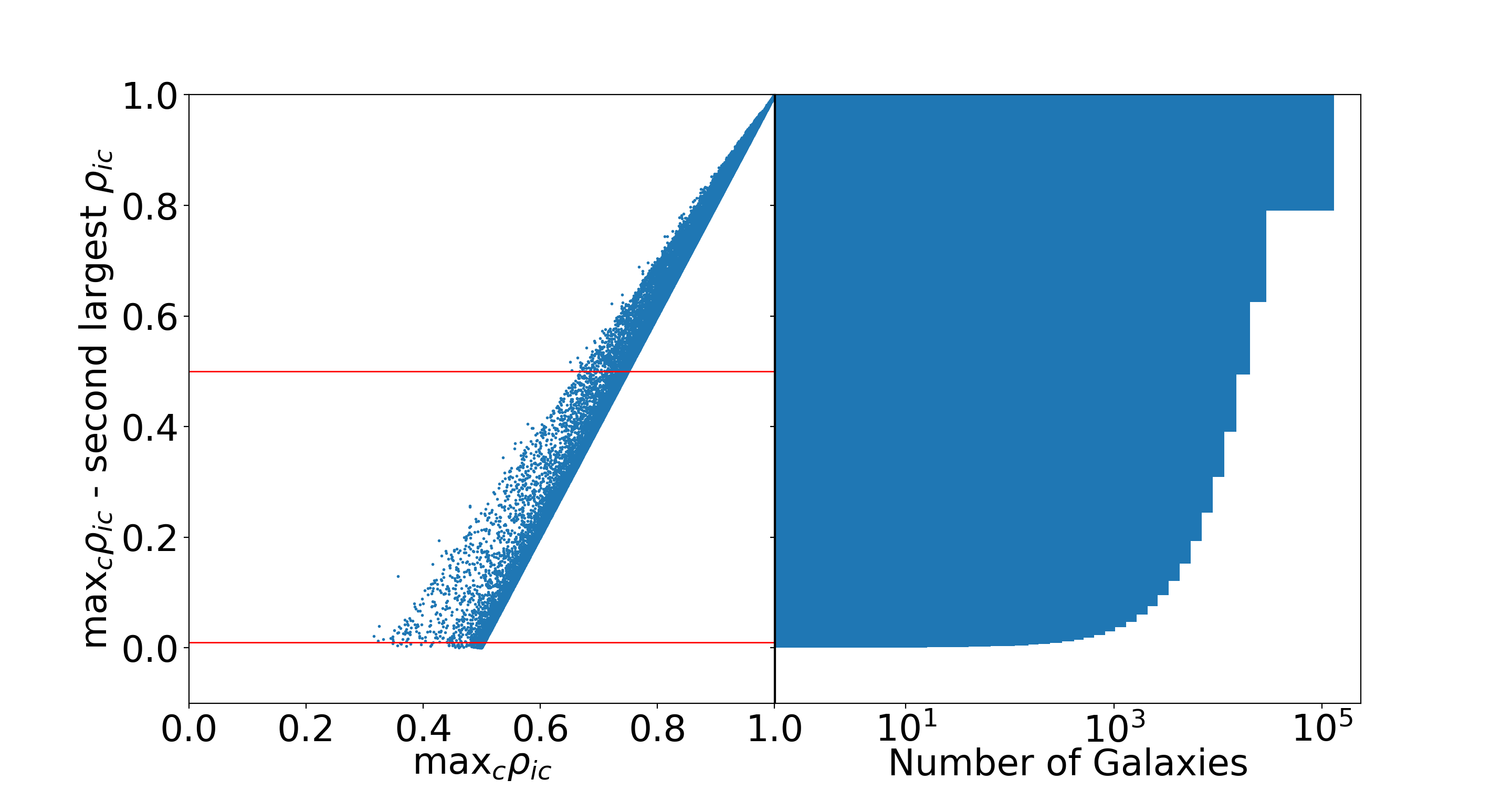}
\caption{The difference between the SoDDA probabilities of the most likely and second most likely class for each galaxy in the SDSS D8 sample. The difference is plotted against the probability of the most likely class. The red lines corresponds to a difference of $1\%$ and $50\%$.  Only 0.21\% of the galaxies exhibit a difference between the probabilities of the most and second most likely classes of less than 1\%.  $87.8\%$ of the galaxies have  $\max_c\rho_{ic} > 75\%$, indicating a highly confident classification.  The histogram in the right of the plot shows the cumulative distribution of the difference between the maximum and the second highest probability. It is clear that more than 75\% of the galaxies have difference well above 0.8.}
\label{fig:comment6}
\end{figure*}

 SoDDA provides a robust classification for the vast majority of the galaxies in the SDSS DR8 sample.  For $87.8\%$ of the galaxies, $\max_c\rho_{ic}$ is greater than 75\%. That is, the most probable class for each of $87.8\%$ of the galaxies has a posterior probability greater than 75\%, indicating  strong  confidence in the  adopted classification (the difference in the classification probability with the second largest class is at least 50\%). 
{\new{ 
The difference between the largest and the second largest $\rho_{ic}$ (among the classes for each object), is a good indicator of the uncertainty of the classification. We find that this difference is greater than $50\%$ for $88.3\%$ of the galaxies, suggesting that the classifications are robust for the vast majority of the sample. The difference between the $\max_c\rho_{ic}$  and the second largest $\rho_{ic}$ is smaller than $10\%$ for $2.1\%$ of the galaxies, and smaller than $1\%$ for only $0.17\%$ of the galaxies. This  indicates that the classification is uncertain for very few galaxies in the overall sample. 
}}
 This is illustrated in Figure~\ref{fig:comment6}  which plots  $\max_c \rho_{ic}$, against the difference between $\max_c\rho_{ic}$  and the second largest $\rho_{ic}$ among the classes. The red lines denote a difference between the two highest  values of $\rho_{ic}$ (among the classes) of $1\%$ and $50\%$. 

\begin{figure*}
\centering
\includegraphics[width=0.9\linewidth]{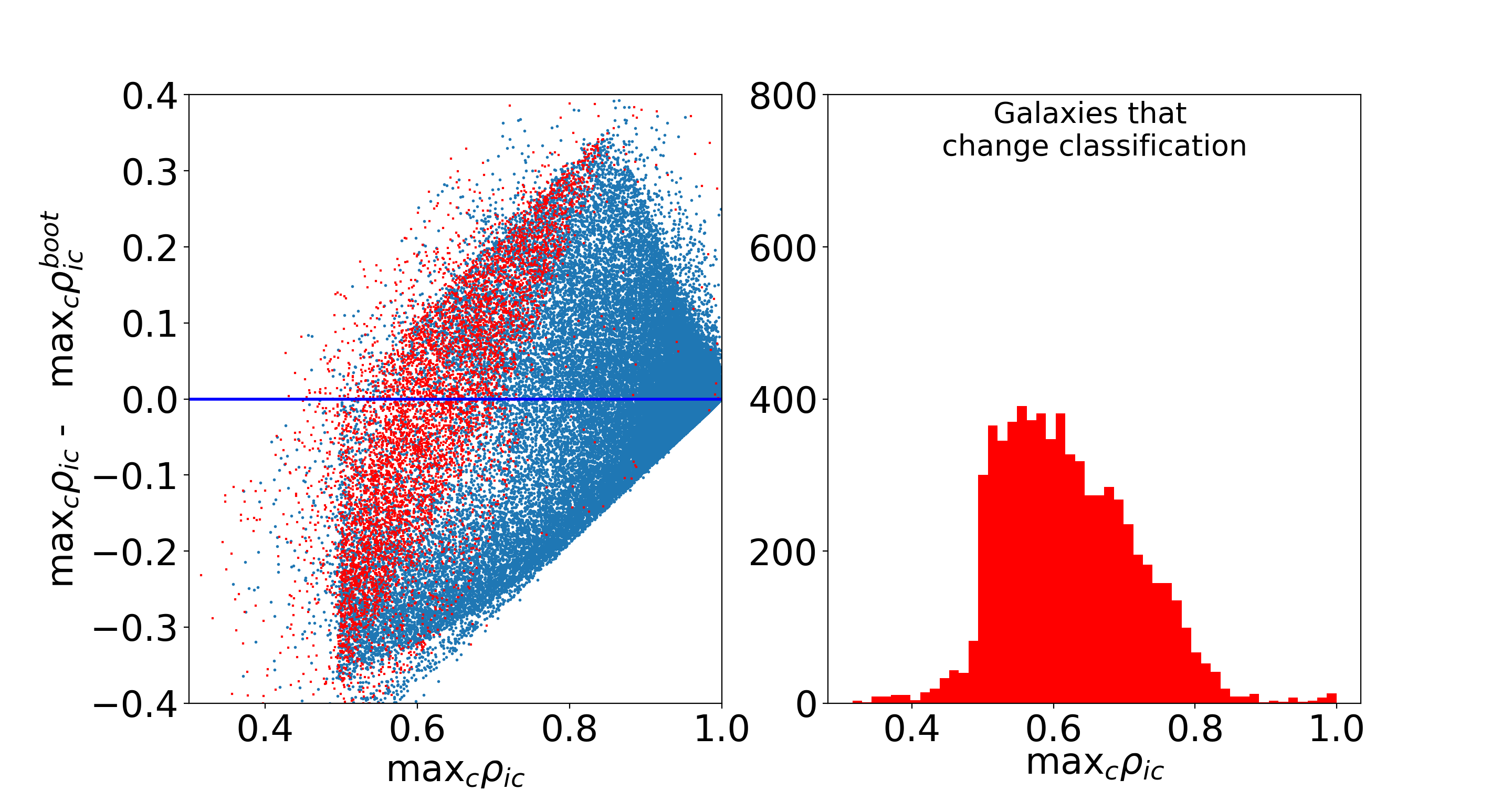}
\caption{(left)A plot of the difference between the class probabilities of the individual galaxies computed with the full data and with the bootstrap sample, namely a plot of $\max_c \rho_{ic} - \max_c \rho_{ic}^{\rm boot}$ against $\max_c \rho_{ic}$. Galaxies that are classified differently by the two schemes are plotted in red. The vast majority of galaxies have the same classification under both schemes; those that do not (5.1\% of the full sample) have $\max_c\rho_{ic} < 75\%$ (89.9\% of them),   meaning their classification was not clear to begin with.  (right) A histogram of the distribution of the maximum probability (i.e. the probability of the highest class $\max_c \rho_{ic}$) for the objects that change classification in the bootstrap analysis. {\newtwo{The vast majority of the objects have  $\max_c \rho_{ic}<0.75$}}. Note the sheer difference in the number of objects that change classification with respect to the total number of objects.}
\label{fig:bootstrap}
\end{figure*}

{\VS In order to assess the stability of the classification we randomly select a bootstrap sample consisting of  90\% of the SDSS DR8 data (sampled without replacement). Using the bootstrap sample, we retune the classifier by estimating the means, weights, and covariance matrices for the {\new 20}  subpopulations, assigning each to one of the 4 activity classes, and recalculating the probability that each galaxy  belongs to each of the 4 classes.  We denote these probabilities, $\rho_{ic}^{\rm boot}$, to distinguish them from those computed with the full SDSS DR8 sample, namely $\rho_{ic}$. There is excellent agreement between the original classification and that obtained using the bootstrap sample. Specifically, {\new{94.9\%}} of the galaxies  are classified into the same activity type with both classifiers. {\new{Similarly, 88.4\% of the galaxies classified as Composites (the class with the largest degree of mixing with the other classes; c.f. Figs. \ref{fig:class_pred}, \ref{fig:comment7}) using  the original classifier are classified in the same way using the set of parameters obtained from using the bootstrap sample. The figures are 95.1\% for Seyferts, 98.9\% for LINERs, and 95.8\% for SFGs.}}

Overall there is little difference between the class probabilities of the individual galaxies computed with the full data and with the bootstrap sample. To illustrate this, we plot  $\max_c \rho_{ic} - \max_c \rho_{ic}^{\rm boot}$ against $\max_c \rho_{ic}$
in Figure~\ref{fig:bootstrap}. Galaxies that are classified differently by the two classifiers are plotted in red.  Again, there is excellent agreement: Not only is the classification of the vast majority of galaxies  the same for both classifiers, but the probabilities of belonging to the chosen class are both similar and high. Of the  galaxies (5.1\%) that are classified differently, 89.9\% have $\max_c \rho_{ic} < 75\%$,  meaning their classification was not clear to begin with. Overall, our classifier appears robust to the choice of sample used for defining the classification clusters.}

\section{Comparison with 2-dimensional Classification Scheme}

In contrast to 
 {\AZ the standard approach of using hard thresholds to define the different classes}, SoDDA uses soft clustering. {\new{ This allows for the natural mixing between the different classes given that there is a continuous distribution of galaxies in the emission-line diagnostic diagrams.}}   We thus calculate the posterior probability of each galaxy belonging to each activity class. Moreover,  SoDDA is not based on any particular set of {\AZ two-dimensional} projections of the distributions of emission-line ratios, but rather it takes into account the joint distribution of all 4 emission-line ratios, {\new{which maximizes the discriminating power of the diagnostic.}} Thus, the main difference between the two schemes is  that  SoDDA does not produce {\VS contradictory classifications for the same galaxy.
Rather} SoDDA provides a single coherent summary {\AZ based on {\textit{all}} diagnostic line ratios}:  a posterior membership probability for each galaxy.{\new{ This allows us to select a sample of galaxies at the desired level of confidence, either in terms of absolute probability of belonging in a given class, or in terms of the odds in belonging in different classes. }}

\begin{landscape}
\begin{table}
\caption{ A 3-way classification table that compares the SoDDA classification with the standard, 2-dimensional classification scheme \citep{kewley2006host}. Each cell has 3 values: the number of galaxies with (i) $\rho_{ic}\geq 75\%$, (ii)   $50\%\leq \rho_{ic}<75\%$, and (iii)  $\rho_{ic}<50\%$, where $\rho_{ic}$ is the posterior probability that galaxy $i$ belongs to galaxy class $c$ under SoDDA. {\VS Contradictory classifications are called ambiguous classifications by \citet{kewley2006host}}.}

\tabcolsep=0.11cm
\begin{tabular}{c c r r r r r r r r r r r r r r r r r r r  r r r r  }
\multirow{7}{*}{\rotatebox[origin=c]{90} {SoDDA}}  & \multicolumn{24}{c}{Kewley et al. (2006) } \\
              & & \multicolumn{3}{c}{SFGs} &&  \multicolumn{3}{c}{Seyferts} &&  \multicolumn{3}{c}{LINERs} & & \multicolumn{3}{c}{Comp} &&  \multicolumn{3}{c}{Contradictory}  &&  \multicolumn{3}{c}{Total} \\

 &  & $\geq 75\%$ & $50\%-75\%$ & $<50\%$ &   & $\geq 75\%$ & $50\%-75\%$ & $<50\%$     & &      $\geq 75\%$ & $50\%-75\%$ & $<50\%$       &  &   $\geq 75\%$ & $50\%-75\%$ & $<50\%$      && $\geq 75\%$ & $50\%-75\%$ & $<50\%$       &&    $\geq 75\%$ & $50\%-75\%$ & $<50\%$  \\
 \cmidrule{3-5}\cmidrule{7-9}\cmidrule{11-13} \cmidrule{15-17} \cmidrule{19-21} \cmidrule{23-25} \\
 & SFGs & 98363 & 3521 & 42 &   & 4 & 2 &0     & &      0& 1 & 0       &  &   1535 & 2369 & 113      && 1745 & 99 & 13       &&    101647 &  5992 &168  \\
&Seyferts            & 0 & 1 & 1 & &   3462 & 241 & 1     & &      30& 48 &7         &&   80 & 336 & 42      && 532 & 497 & 45       & &   4104 &  1123 &96  \\
&LINERs          & 0 & 0 & 0&   & 0&0& 0      &&      811& 354 & 21         &&   436 &255& 23      && 34 & 44 & 26       &&    1281 &  653 &70 \\
&Comp     & 43 & 791 & 38& &    0 &0 & 0 &&      21&147& 24         &&   7545 & 6438 & 207     & & 157 & 208 & 46      & &    7766 & 7584 &315  \\

\end{tabular}
\label{tab:3way}
\end{table}
\end{landscape}

A 3-way classification table that compares SoDDA with the {\AZ commonly used} scheme proposed by \citet{kewley2006host} appears in Table~\ref{tab:3way}. Each cell has 3 values: the number of galaxies with (i) $\rho_{ic}\geq 75\%$, (ii)   $50\%\leq \rho_{ic}<75\%$, and (iii)  $\rho_{ic}<50\%$, where $\rho_{ic}$ is the posterior probability that galaxy $i$ belongs to galaxy class $c$.  For example, the cell in the first row and first column shows that of the galaxies that both the SoDDA  and the \citet{kewley2006host} method classify as SFG, {\new{98,363}}  are SFGs under SoDDA with probability greater than $75\%$, {\new{3,521}} with probability between $50\%$ and $75\%$, and  only {\new{42}} with probability less than $50\%$.
{\new{In general there is very  good agreement between the SoDDA and the \citet{kewley2006host} classification for the star-forming and the Seyfert galaxy classes. In the case of LINERs there is also reasonable agreement, but with a larger fraction of galaxies classified in the intermediate confidence ($50\%\leq \rho_{ic}<75\%$) regime. In the case of composite objects, however, the fraction of galaxies classified in the intermediate or low ($\rho_{ic}<50\%$) confidence regime increases dramatically. This is a result of the overlap between the composite and the other activity classes in the $\log($[\ion{N}{II}]$/$H$\alpha$), $\log$([\ion{S}{II}]$/$H$\alpha$), and  $\log$([\ion{O}{III}]$/$H$\beta$), but not for  ([\ion{O}{I}]$/H\alpha$ - [\ion{O}{III}]$/H\beta$) and the ([\ion{S}{II}]/H$\alpha$)- [\ion{O}{III}]$/H\beta$) diagnostics  (Fig. \ref{fig:class_pred}, \ref{fig:SFG}).    }}
{\new{The majority of the galaxies  that have contradictory classifications according to \citet{kewley2006host} are estimated with the SoDDA to be SFGs, and increasingly reduced fractions are allocated to the Seyfert, LINER, and Composite classes. }}

{\new{ In Figure~\ref{fig:class_pred} we show the classification based on the diagnostic lines presented in \citet{kewley2006host} (top panels) along with the classification based on the SoDDA method. The colour coding of the different classes is the same in both panels (red for SFGs, yellow for Seyfert, blue for LINERs, green for composite galaxies). Objects with contradictory classifications in the top panel are marked in black.}} 
 {\VS  The overlap between the composite galaxies (green) and the SFGs (red) is clear in the SoDDA classification (middle and right panels of Figure~\ref{fig:class_pred}),} indicating that the 2-dimensional projection of this 4-dimensional parameter space is insufficient  for capturing its complex structure and accurately classifying the galactic activity. {\AZ The use of hard boundaries defined independently in the 2-dimensional projections is responsible for those galaxies with  contradictory classification. On the other hand the probabilistic approach of SoDDA simultaneously accounts for the 4-dimensional structure of the data space and inherently alleviates these inconsistent classifications, while at the same time giving a confident classification of the galaxies to activity classes.}

\section{Multidimensional Decision Boundaries}

 {\AZ In order to provide a more immediately usable diagnostic in the spirit of the classification lines of \citet{kauffmann2003host} and \citet{kewley2006host}, which however,  {\textit{simultaneously}} employ  the information in all diagnostic lines, we use a support vector machine (SVM) \citep{cortes1995support} to obtain multidimensional decision boundaries based on the SoDDA results.}   A SVM is a discriminative classifier formally defined by a separating hyperplane. In other words, given classified galaxies, the algorithm outputs an optimal hyperplane which can be used to categorize new unlabelled galaxies. {\new{This hybrid approach uses the SoDDA classification to disentangle the complex multi-dimensional structure of the overlapping clusters, while providing easy to use diagnostic surfaces in the spirit of the commonly BPT-like diagnostics.}}

\subsection{4-dimensional Decision Boundaries}
The input data {\AZ for the derivation of the multidimensional decision boundaries}  are the 4 emission line ratios for the galaxies in the {\VS SDSS DR8 sample} (i.e. $x$), and the classification for each galaxy {\AZ as} obtained with SoDDA {\newtwo{(i.e.,  $y$)}}. We use the {\tt scikit-learn} Python library to fit the SVM model,  employing a linear kernel function. {\AZ A more complex function did not provide an improvement significant enough to justify its use, {\new{ especially}} given the simplicity of a linear kernel.} The SVM algorithm requires tuning the cost factor parameter $C$, that sets the width of the margin between hyperplanes separating different classes of objects. After a grid search in a range of values for $C$, we {\new{adopt}} a value of $C=1$ based on 10-fold cross-validation. $\cal K$-fold cross-validation is a model validation method for estimating the performance of the model. The data is split in $\cal K$ roughly equal parts. For each $ \kappa \in (1,..., \cal K)$ we fit the model in the other $\cal {K} $$-1$ parts of the data and calculate the prediction error of the fitted model when predicting the $\kappa$th part of the data (the error is effectively the number of inconsistent classifications between the SVM analysis on the the $\kappa$th part of the data and the classifications obtained by SoDDA for the same galaxies). By repeating this procedure for a range of values for the model parameters ($C$), we choose the values of $C$ that give us the SVM model with the minimum expected prediction error.

Using the SoDDA classification, we employ an SVM approach  to define multidimensional surfaces separating the galaxy activity classes. More specifically, we find an optimal separation hyperplane  using  the 4 emission line ratios for the  galaxies  from {\VS the SDSS DR8 sample} and  their most probable classification obtained by SoDDA as inputs. The 4-dimensional linear decision boundaries for the four galaxy classes are defined below.

\noindent \textbf{SFG:}
%
%
\begin{align}
-7.31  \log( [ \text{\ion{N}{II}}]  / \text{H}\alpha) &+2.75 \log( [ \text{\ion{S}{II}}]  / \text{H}\alpha) -1.41\log(  \text{[\ion{O}{I}]} / \text{H}\alpha) \nonumber \\
&-5.91 \log(  \text{[\ion{O}{III}]} / \text{H}\beta) >1.92\\
-5.32 \log( [ \text{\ion{N}{II}}]  / \text{H}\alpha)&-6.37  \log( [ \text{\ion{S}{II}}]  / \text{H}\alpha) -3.40 \log(  \text{[\ion{O}{I}]} / \text{H}\alpha)  \nonumber \\
&-0.42 \log(  \text{[\ion{O}{III}]} / \text{H}\beta) >6.51 \\
-23.01  \log( [ \text{\ion{N}{II}}]  / \text{H}\alpha) &+0.93  \log( [ \text{\ion{S}{II}}]  / \text{H}\alpha) -5.30 \log(  \text{[\ion{O}{I}]} / \text{H}\alpha)  \nonumber  \\
& -8.10  \log(  \text{[\ion{O}{III}]} / \text{H}\beta) >16.38
\end{align}


\noindent \textbf{Seyferts:}
%
\begin{align}
-7.31  \log( [ \text{\ion{N}{II}}]  / \text{H}\alpha) &+2.75 \log( [ \text{\ion{S}{II}}]  / \text{H}\alpha) -1.41\log(  \text{[\ion{O}{I}]} / \text{H}\alpha) \nonumber \\
&-5.91 \log(  \text{[\ion{O}{III}]} / \text{H}\beta) <1.92\\
0.37  \log( [ \text{\ion{N}{II}}]  / \text{H}\alpha)& -4.55  \log( [ \text{\ion{S}{II}}]  / \text{H}\alpha)
 -7.21 \log(  \text{[\ion{O}{I}]} / \text{H}\alpha) ) \nonumber \\
 &+11.65  \log(  \text{[\ion{O}{III}]} / \text{H}\beta) >10.02 \\
7.14 \log( [ \text{\ion{N}{II}}]  / \text{H}\alpha) &-3.12  \log( [ \text{\ion{S}{II}}]  / \text{H}\alpha) 
+0.46 \log(  \text{[\ion{O}{I}]} / \text{H}\alpha)  \nonumber \\
&+16.08 \log(  \text{[\ion{O}{III}]} / \text{H}\beta) >2.82
\end{align}

\noindent \textbf{LINERs:}
%
\begin{align}
-5.32 \log( [ \text{\ion{N}{II}}]  / \text{H}\alpha)&-6.37 \log( [ \text{\ion{S}{II}}]  / \text{H}\alpha) -3.40 \log(  \text{[\ion{O}{I}]} / \text{H}\alpha)  \nonumber \\
&-0.42\log(  \text{[\ion{O}{III}]} / \text{H}\beta) <6.51 \\
0.37  \log( [ \text{\ion{N}{II}}]  / \text{H}\alpha)& -4.55  \log( [ \text{\ion{S}{II}}]  / \text{H}\alpha)
 -7.21 \log(  \text{[\ion{O}{I}]} / \text{H}\alpha)  \nonumber \\
 &+11.65  \log(  \text{[\ion{O}{III}]} / \text{H}\beta) <10.02 \\
-1.04  \log( [ \text{\ion{N}{II}}]  / \text{H}\alpha) &+8.94  \log( [ \text{\ion{S}{II}}]  / \text{H}\alpha)
+6.48\log(  \text{[\ion{O}{I}]} / \text{H}\alpha)  \nonumber  \\
&+6.69  \log(  \text{[\ion{O}{III}]} / \text{H}\beta) >-6.90
\end{align}

\noindent \textbf{Composites:}
%
\begin{align}
-23.01  \log( [ \text{\ion{N}{II}}]  / \text{H}\alpha) &+0.93  \log( [ \text{\ion{S}{II}}]  / \text{H}\alpha) -5.30 \log(  \text{[\ion{O}{I}]} / \text{H}\alpha)  \nonumber  \\
& -8.10  \log(  \text{[\ion{O}{III}]} / \text{H}\beta) <16.38\\
7.14 \log( [ \text{\ion{N}{II}}]  / \text{H}\alpha) &-3.12  \log( [ \text{\ion{S}{II}}]  / \text{H}\alpha) 
+0.46 \log(  \text{[\ion{O}{I}]} / \text{H}\alpha)  \nonumber \\
&+16.08 \log(  \text{[\ion{O}{III}]} / \text{H}\beta) <2.82\\
-1.04  \log( [ \text{\ion{N}{II}}]  / \text{H}\alpha) &+8.94  \log( [ \text{\ion{S}{II}}]  / \text{H}\alpha)
+6.48\log(  \text{[\ion{O}{I}]} / \text{H}\alpha)  \nonumber  \\
&+4.69  \log(  \text{[\ion{O}{III}]} / \text{H}\beta) <-6.90
\end{align}

Table~\ref{tab:svm_data} compares the SoDDA classification  with the proposed classification from the SVM, while Table~\ref{tab:svm_kewley} compares the scheme from \citet{kewley2006host} with the SVM. We see excellent agreement between the SoDDA and the SVM-based classification. {\VS More specifically, 99.0\% of the galaxies classified as SFGs by SoDDA are classified in the same way as the SVM-based classification. The figures are 96.9\% for Seyferts, 91.2\% for LINERs, and 90.2\% for Composites.}
{\new{Similarly, we find very good agreement between the traditional 2-dimensional diagnostics of \citep{kewley2006host} and the SVM method in the cases of SFGs and Seyfert galaxies (Table \ref{tab:svm_kewley}). For Composite objects and LINERs we find a larger number of objects for which we obtain a different classification based on the two methods. The largest discrepancy is in the case of LINERs (agreement for 80\% of the LINER sample), which we attribute to the complex shape on the distribution of the Composite objects for which the SoDDA analysis shows that they extend to the locus of LINERs  (Figs. \ref{fig:class_pred}, \ref{fig:SFG}).  We note that such discrepancies are expected, given the ad-hoc definition of the activity classes, particularly in the case of composite galaxies.  }}

\begin{table*}
\caption{{\AZ Comparison of } the SoDDA classification with that of the 4-dimensional SVM ([\ion{O}{III}]$/$H$\beta$,  [\ion{N}{II}]$/$H$\alpha$, [\ion{S}{II}]$/$H$\alpha$ and [\ion{O}{I}]$/$H$\alpha$ space).}
\begin{tabular}{c c c c c c c   }
\multirow{7}{*}{\rotatebox[origin=c]{90} {SVM}}  & \multicolumn{6}{c}{SoDDA } \\
              & & SFGs & Seyferts & LINERs & Composites   & Total \\
\cline{2-7}
 & SFGs & 106782 &     14      &       13         &   1330      & 108139  \\
&Seyferts            & 36 &    5157       &      39       &   115       &   5347  \\
&LINERs          & 22&    9      &      1828     &   85      & 1944 \\
&Composites     & 967 &    143     &      124         &   14135    &  15369 \\
&Total    & 107807&   5323     &      2004        &   15665   &   \\

\end{tabular}
\label{tab:svm_data}
\end{table*}

\begin{table*}
\caption{{\AZ Comparison of} the classifications of a 4-dimensional SVM with that of the method by \citet{kewley2006host} ([\ion{O}{III}]$/$H$\beta$,  [\ion{N}{II}]$/$H$\alpha$, [\ion{S}{II}]$/$H$\alpha$ and [\ion{O}{I}]$/$H$\alpha$ space).  {\VS Contradictory classifications are called ambiguous classifications by \citet{kewley2006host}}.}
\begin{tabular}{c c c c c c c   c}
\multirow{7}{*}{\rotatebox[origin=c]{90} {SVM}}  & \multicolumn{7}{c}{\citet{kewley2006host} } \\
              & & SFGs & Seyferts & LINERs & Composites   &Contradictory & Total \\
\cline{2-8}
 & SFGs & 102455 &     0      &       0         &   3987      &1697 & 108139  \\
&Seyferts            & 0 &    3708       &      107       &   478&1054       &   5347  \\
&LINERs          & 0&    0      &      1176    &   677& 91      & 1944 \\
&Composites     & 345 &    2     &     181       &   14237    &604 &  15369 \\
&Total         & 102800&   3710     &      1464  &19379      &   3446   &  \\

\end{tabular}
\label{tab:svm_kewley}
\end{table*}

\subsection{3-dimensional Decision Boundaries}
\label{sec:3d_boundaries}
Because the [\ion{O}{I}] line is generally {\AZ very weak and hence} hard to {\AZ measure}, it is common to {\AZ use the flux ratios of the five other strong lines in the optical spectrum:} $\log($[\ion{N}{II}]$/$H$\alpha$), $\log$([\ion{S}{II}]$/$H$\alpha$), and  $\log$([\ion{O}{III}]$/$H$\beta$). 
Thus, we {\new{ use the SoDDA classification (Section~\ref{dataset}) as the basis for the definition of decision boundaries by applying the SVM algorithm in the 3-dimensional space defined by the ($\log($[\ion{N}{II}]$/$H$\alpha$), $\log$([\ion{S}{II}]$/$H$\alpha$), and  $\log$([\ion{O}{III}]$/$H$\beta$) emission-line ratios.}}   The resulting 3-dimensional decision surfaces for the four galaxy classes are presented below. 



\noindent \textbf{SFG:}
\begin{align}
-7.27  \log( [ \text{\ion{N}{II}}]  / \text{H}\alpha) &+1.523 \log( [ \text{\ion{S}{II}}]  / \text{H}\alpha)  \nonumber \\
&-7.02 \log(  \text{[\ion{O}{III}]} / \text{H}\beta) >0.25\\
-4.08 \log( [ \text{\ion{N}{II}}]  / \text{H}\alpha)&-9.33  \log( [ \text{\ion{S}{II}}]  / \text{H}\alpha)  \nonumber \\
&-1.93 \log(  \text{[\ion{O}{III}]} / \text{H}\beta) >3.28 \\
-19.55  \log( [ \text{\ion{N}{II}}]  / \text{H}\alpha) &-3.07  \log( [ \text{\ion{S}{II}}]  / \text{H}\alpha)   \nonumber  \\
& -7.10  \log(  \text{[\ion{O}{III}]} / \text{H}\beta) >9.45
\end{align}


\noindent \textbf{Seyferts:}
\begin{align}
-7.27  \log( [ \text{\ion{N}{II}}]  / \text{H}\alpha) &+1.523 \log( [ \text{\ion{S}{II}}]  / \text{H}\alpha)  \nonumber \\
&-7.02 \log(  \text{[\ion{O}{III}]} / \text{H}\beta) <0.25\\
0.23 \log( [ \text{\ion{N}{II}}]  / \text{H}\alpha)&-9.66  \log( [ \text{\ion{S}{II}}]  / \text{H}\alpha)  \nonumber \\
&+9.29 \log(  \text{[\ion{O}{III}]} / \text{H}\beta) >4.03 \\
7.22  \log( [ \text{\ion{N}{II}}]  / \text{H}\alpha) &-2.77 \log( [ \text{\ion{S}{II}}]  / \text{H}\alpha)   \nonumber  \\
& +16.04  \log(  \text{[\ion{O}{III}]} / \text{H}\beta) >3.23
\end{align}

\noindent \textbf{LINERs:}
\begin{align}
-4.08 \log( [ \text{\ion{N}{II}}]  / \text{H}\alpha)&-9.33  \log( [ \text{\ion{S}{II}}]  / \text{H}\alpha)  \nonumber \\
&-1.92 \log(  \text{[\ion{O}{III}]} / \text{H}\beta) <3.28 \\
0.23 \log( [ \text{\ion{N}{II}}]  / \text{H}\alpha)&-9.66  \log( [ \text{\ion{S}{II}}]  / \text{H}\alpha)  \nonumber \\
&+9.29 \log(  \text{[\ion{O}{III}]} / \text{H}\beta) <4.03 \\
-0.13 \log( [ \text{\ion{N}{II}}]  / \text{H}\alpha)&+13.16  \log( [ \text{\ion{S}{II}}]  / \text{H}\alpha)  \nonumber \\
&+5.04 \log(  \text{[\ion{O}{III}]} / \text{H}\beta) >-1.84 
\end{align}

\noindent \textbf{Composites:}
\begin{align}
-19.55  \log( [ \text{\ion{N}{II}}]  / \text{H}\alpha) &-3.07  \log( [ \text{\ion{S}{II}}]  / \text{H}\alpha)   \nonumber  \\
& -7.10  \log(  \text{[\ion{O}{III}]} / \text{H}\beta) <9.45 \\
7.22  \log( [ \text{\ion{N}{II}}]  / \text{H}\alpha) &-2.77  \log( [ \text{\ion{S}{II}}]  / \text{H}\alpha)   \nonumber  \\
& +16.04  \log(  \text{[\ion{O}{III}]} / \text{H}\beta) <3.23\\
-0.13 \log( [ \text{\ion{N}{II}}]  / \text{H}\alpha)&+13.16  \log( [ \text{\ion{S}{II}}]  / \text{H}\alpha)  \nonumber \\
&+5.04 \log(  \text{[\ion{O}{III}]} / \text{H}\beta) <-1.84 
\end{align}

The multidimensional decision boundaries  achieve a mean classification accuracy of about $96.7\%$ based on 10-fold cross validation {\VS with respect to the SoDDA classification}. Table~\ref{tab:svm3d_data} compares the SoDDA classification  with the proposed classification from the SVM, while Table~\ref{tab:svm3d_kewley} compares the scheme from \citet{kewley2006host} with the SVM. As with the 4-dimensional SVM classification, we have excellent agreement with the SoDDA classification and slightly worse agreement with the traditional 2-dimensional diagnostics. Surprisingly, we also find very {\new{good}} agreement between the 3-dimensional and the 4-dimensional SVM diagnostics indicating that removing the fourth line ratio ([\ion{O}{I}]$/$H$\alpha$) does not significantly affect the quality of the classification.  {\VS More specifically, 98.7\% of the galaxies classified as SFGs by SoDDA are classified in the same way by the 3-dimensional SVM-based classification. The figures are 96.1\% for Seyferts, 76.0\% for LINERs, and 85.4\% for Composites.} 
In other words, removing the ([\ion{O}{I}]$/$H$\alpha$) line ratio has no impact on the classification error for SFGs and the Seyferts, and {\AZ results in a different classification of   $10.9\%$ of galaxies classified as LINERs by SoDDA and $3.7\%$ of galaxies classified as Composites by SoDDA, when compared to the complete 4-dimensional diagnostic}. 

\begin{table*}
\caption{{\AZ Comparison of} the classifications of SoDDA with that of the 3-dimensional SVM ([\ion{O}{III}]$/$H$\beta$,  [\ion{N}{II}]$/$H$\alpha$, and [\ion{S}{II}]$/$H$\alpha$ space.}
\begin{tabular}{c c c c c c c   }
\multirow{7}{*}{\rotatebox[origin=c]{90} {SVM}}  & \multicolumn{6}{c}{SoDDA } \\
              & & SFGs & Seyferts & LINERs & Composites   & Total \\
\cline{2-7}
 & SFGs & 106416 &     16      &       27         &   1965     & 108424  \\
&Seyferts      & 40 &    5117       &      111       &   108       &   5376  \\
&LINERs       & 31&    68      &      1524     &   217      & 1840 \\
&Composites   & 1320 &    122     &      342         &   13375    &  15159 \\
&Total    & 107807&   5323    &      2004        &   15665   &   \\

\end{tabular}
\label{tab:svm3d_data}
\end{table*}

\begin{table*}
\caption{{\AZ Comparison of} the classifications of the 3-dimensional SVM and that of the method by \citet{kewley2006host} ([\ion{O}{III}]$/$H$\beta$,  [\ion{N}{II}]$/$H$\alpha$, [\ion{S}{II}]$/$H$\alpha$ and [\ion{O}{I}]$/$H$\alpha$ space). {\VS Contradictory classifications are called ambiguous classifications by \citet{kewley2006host}}.}
\begin{tabular}{c c c c c c c   c}
\multirow{7}{*}{\rotatebox[origin=c]{90} {SVM}}  & \multicolumn{7}{c}{\citet{kewley2006host} } \\
              & & SFGs & Seyferts & LINERs & Composites   &Contradictory & Total \\
\cline{2-8}
 & SFGs & 102750 &     0      &       0         &   3777      &1897 & 108424  \\
&Seyferts            & 0 &    3708       &      173       &   490 &1005       &   5376  \\
&LINERs          & 0&    0      &      1101    &   601&138      & 1840 \\
&Composites     & 50 &    2     &     190       &   14511    &406 &  15159 \\
&Total    & 102800&   3710     &      1464  &19379      &   3446   &   \\

\end{tabular}
\label{tab:svm3d_kewley}
\end{table*}

\section{Discussion}

We propose a  new soft clustering scheme, the Soft Data-Driven Allocation (SoDDA) method, for classifying galaxies using emission-line ratios. Our method uses {\AZ an optimal} number of MG subpopulations in order to capture the multi-dimensional structure of the dataset and afterwards concatenate the MG subpopulations into clusters by assigning them to different activity types, based on the location of their means with respect to the loci of the activity classes as defined by \citet{kewley2006host}. 

 The main advantages of this method are{\new{: (a)}} the use of all four optical-line ratios simultaneously, thus maximising the available information{\new{, }}avoiding contradicting classifications, and {\new{ (b)}} treating each class as a distribution resulting in soft classification boundaries. {\new{This allows us to account for the inherent overlap between the different activity classes stemming from the simultaneous presence of different excitation mechanisms with a varying degree of intensity.  We achieve this by calculating the probability for an object to be associated with each one of these activity classes given their distribution in the multi-dimensional diagnostic space.  }}

An issue with data-driven classification {\new{methods}} is the question of whether the data have sufficient discriminating power to distinguish the different activity classes. A strong indication in this direction comes from the fact that the original BPT diagnostic  \citep{baldwin1981classification}  and its more recent redefinition by \citet{kauffmann2003host} and \citet{kewley2006host} was driven by the clustering of the activity classes in different loci on the 2-dimensional line-ratio diagrams. Furthermore, this distinction was supported by photoionisation models  \citep{kewley2001theoretical,2013ApJ...774L..10K}  which indicate  that while there is a continuous evolution of the location of sources on the  2-dimensional diagnostic diagrams as a function of their metallicity and hardness of the ionising continuum, star-forming galaxies occupy a distinct region of this diagram.       
In our analysis we follow a hybrid approach in which we identify clusters based on the multi-dimensional distribution of the object line-ratios, and we associate the clusters with activity types based on their location in the standard 2-dimensional diagnostic diagrams. This gives a physical {\AZ interpretation} to each cluster, while tracing the multi-dimensional distribution of {\new{their}} line ratios.

The approach followed in this paper treats the multi-dimensional emission-line diagnostic diagram as a mixture of different classes. This is a more realistic approach as it does not assume fixed boundaries between the activity classes.  Instead, it takes into account {\AZ the fact} that the emission-line ratios of the different activity classes may overlap, {\AZ which is reflected on the probabilities for an object to belong to a given class}. This in fact is reflected in the often inconsistent classification between different 2-dimensional diagnostics \citep{ho1997search,2010ApJ...709..884Y}, and  is clearly seen in the complex structure of the locus of the activity classes in {\AZ the} 3-dimensional rotating diagnostics available {\AZ in} the online supplements. Therefore, the optimal way to characterize a galaxy is by calculating {\new{the\textit{ probability}}} that it {\AZ belongs} to each of the activity classes, {\new{ instead of associating it unequivocally with a given class.}} {\AZ This also gives us the possibility to define samples of different types of galaxies at various confidence levels.}

 Another advantage of this approach is that we take into account {\AZ \textit{all}} available information for the activity classification of galactic nuclei. This is important given the complex shape of the multi-dimensional distributions of the emission line ratios  (e.g. online 3-dimensional rotating diagnostics;  see also \citealt{vogt2014galaxy}). This way we increase the power of the 2-dimensional diagnostic tools, and eliminate the contradicting classifications they often give. 
  This is demonstrated by the excellent agreement between the classification of the 4-dimensional diagnostic ([\ion{O}{III}]/H$\beta$, [\ion{O}{I}]/H$\alpha$, [\ion{N}{II}]/H$\alpha$,  [\ion{S}{II}]/H$\alpha$)  with the 3-dimensional diagnostic excluding the often weak and hard to detect [\ion{O}{I}]  line ([\ion{O}{III}]/H$\beta$, [\ion{O}{I}]/H$\alpha$, [\ion{N}{II}]/H$\alpha$,  [\ion{S}{II}]/H$\alpha$; see \ref{sec:3d_boundaries}). This agreement indicates that the loss of the diagnostic power of the [\ion{O}{I}]/H$\alpha$ line (which {\AZ is considered} the main discriminator between LINERs and other activity classes   \citep[e.g.][]{kewley2006host}) in the 4-dimensional diagnostic, can be compensated by the structure of the locus of the different activity classes which allows their distinction even in the 3-dimensional diagnostic.    

{\newtwo{A very similar approach was followed by \cite{deSouza17}} who modeled the ([\ion{O}{III}]/H$\beta$,  [\ion{N}{II}]/H$\alpha$, EW(H$\alpha$)) 3-dimensional space with a set of 4 multi-dimensional Gaussians. The different number of Gaussian components required in our work is the result of the more complex structure of the distribution of the line ratios in the 4-dimensional ([\ion{O}{III}]/H$\beta$, [\ion{O}{I}]/H$\alpha$, [\ion{N}{II}]/H$\alpha$,  [\ion{S}{II}]/H$\alpha$) space, in comparison to the simpler shape in the 3-dimensional space explored by \cite{deSouza17}. The use of the  EW(H$\alpha$) in the latter study instead of the [\ion{O}{I}]/H$\alpha$ and [\ion{S}{II}]/H$\alpha$ line ratios allow the separation of star-forming from non star-forming galaxies (retired or passive; \cite{cidFernades11}, \cite{Stasinska15}).}

Although the probabilistic clustering contains more information about the classification of each {\newtwo{emission-line}} galaxy, the use of hard decision boundaries for classification is effective and closer to the standard approach used in the  literature. Therefore, we also present hard classification criteria by employing SVM on the distribution of line-ratios of objects assigned to each activity class. The classification accuracy with these hard criteria is $\sim98\%$ when compared to the soft classification (SoDDA).  This indicates that the extended tails of the line-ratio distributions of the different activity classes result in only a small degree of overlap and hence misclassification compared to the results we get from SoDDA. 

{\new{ Several efforts in the past have introduced activity diagnostic tools that combine information from multiple spectral bands and often including spectral-line ratios. For example \cite{2005ApJ...631..163S} and \cite{2012ApJ...748..142D} introduced the use of near and far-IR colours for separating star-forming galaxies from AGN. \citealt{2006ApJ...646..161D} and \citealt{2010ApJ...709.1257T} have further developed the use of IR line diagnostics  (involving for example emission lines from PAHs,  [\ion{O}{IV}], [\ion{Ne}{II}], [\ion{Ne}{III}]), initially proposed by \citealt{SpinoglioMalkan1992}.   }}
 Such diagnostics  have been used extensively in IR surveys in order to address the nature of heavily obscured galaxies, and they are going to be particularly useful for classifying objects detected in surveys performed with the James-Webb Space Telescope. 
 {\new{ Composite diagnostic diagrams involving the [\ion{O}{III}]/H$\beta$ line-ratio and photometric data that are stellar-mass proxies  \cite{Weiner2007}, the stellar mass directly \cite{juneau2011,juneau2014}, or photometric colours \cite{Yan2011}, have been developed to classify high-redshift or heavily obscured objects. In a similar vein, \cite{Stasinska06} propose a diagnostic based on the stellar-population age sensitive $\mathrm{4000\AA}$-break index compared with the equivalent width of the  [\ion{O}{II}]$\mathrm{3727\AA}$ or the [\ion{N}{III}]$\mathrm{3869\AA}$ lines.}}
 
  {\new{ These studies demonstrate that inclusion of information from photometric data, or wavebands other than optical, can extend the use of the diagnostic diagrams to higher redshifts, or increase the sensitivity of the standard diagrams in cases of heavily obscured galaxies or galaxies dominated by old stellar populations. For example, broadening the parameter space to include information from other wavebands (e.g   X-ray luminosity, radio luminosity and spectral index, X-ray to optical flux ratio) along with the multi-dimensional diagnostics discussed in \S5 would further increase the sensitivity of these diagnostic tools by including all available information that would allow us to identify obscured and unobscured AGN, or passive galaxies. The fact that our analysis identifies  multiple subpopulations within each activity class can be used to recognize  subclasses with unusual characteristics that merit special attention. Key for these extensions of the diagnostic tools is to incorporate upper-limits  (i.e., information about the limiting luminosity in a given band in the case of non detections)  and uncertainties in the determination of the clusters in the SoDDA classification or the separating surfaces in the SVM approach. 
}}

\section*{Acknowledgements}

This work was conducted under the auspices of the CHASC International Astrostatistics Center. CHASC is supported by NSF grants DMS 1208791, DMS 1209232, DMS 1513492, DMS 1513484, DMS 1513546, and SI's Competitive Grants Fund 40488100HH0043. We thank CHASC members for many helpful discussions, especially Alexandros Maragkoudakis for providing the data. AZ acknowledges funding from the European Research Council under the European Union's Seventh Framework Programme (FP/2007-2013)/ERC Grant Agreement n. 617001, and support from NASA/ADAP grant NNX12AN05G. This project has been made possible through the ASTROSTAT collaboration, enabled by the Horizon 2020, EU Grant Agreement n. 691164. VLK was supported through NASA Contract NAS8-03060 to the Chandra X-ray Center.

Funding for SDSS-III has been provided by the Alfred P. Sloan Foundation, the Participating Institutions, the National Science Foundation, and the U.S. Department of Energy Office of Science. The SDSS-III web site is http://www.sdss3.org/.

SDSS-III is managed by the Astrophysical Research Consortium for the Participating Institutions of the SDSS-III Collaboration including the University of Arizona, the Brazilian Participation Group, Brookhaven National Laboratory, Carnegie Mellon University, University of Florida, the French Participation Group, the German Participation Group, Harvard University, the Instituto de Astrofisica de Canarias, the Michigan State/Notre Dame/JINA Participation Group, Johns Hopkins University, Lawrence Berkeley National Laboratory, Max Planck Institute for Astrophysics, Max Planck Institute for Extraterrestrial Physics, New Mexico State University, New York University, Ohio State University, Pennsylvania State University, University of Portsmouth, Princeton University, the Spanish Participation Group, University of Tokyo, University of Utah, Vanderbilt University, University of Virginia, University of Washington, and Yale University.




\bibliographystyle{mnras}
\bibliography{references} 


\appendix

\section{{\newtwo{Analysis of the SDSS Data Release 8 sample with SNR$>$3 for [\ion{O}{I}]$/H\alpha$}}}

{\newtwo{The generally used sample for the definition of the activity classification diagnostics in the BPT diagrams employing SDSS data includes the screening criteria listed in \SS3  (e.g. \cite{kewley2006host};  \cite{kauffmann2003host}; \cite{vogt2014galaxy}). However, these screening criteria do not include the generally weak $[\ion{O}{I}]$\,$\mathrm{\lambda6300\AA}$ line. 
In order to assess the sensitivity of our results on the presence of noisy data with low S/N ratio in the $[\ion{O}{I}]$\,$\mathrm{\lambda6300\AA}$ line, we included in the screening criteria presented in \SS3 the criterion of S/N$\geq3$ for the $[\ion{O}{I}]$\,$\mathrm{\lambda6300\AA}$ line. The final sample which has  S/N$\geq3$ in all lines involved, consists of 97,809 galaxies.
}}

\begin{figure*}
\centering
\includegraphics[width=0.75\linewidth]{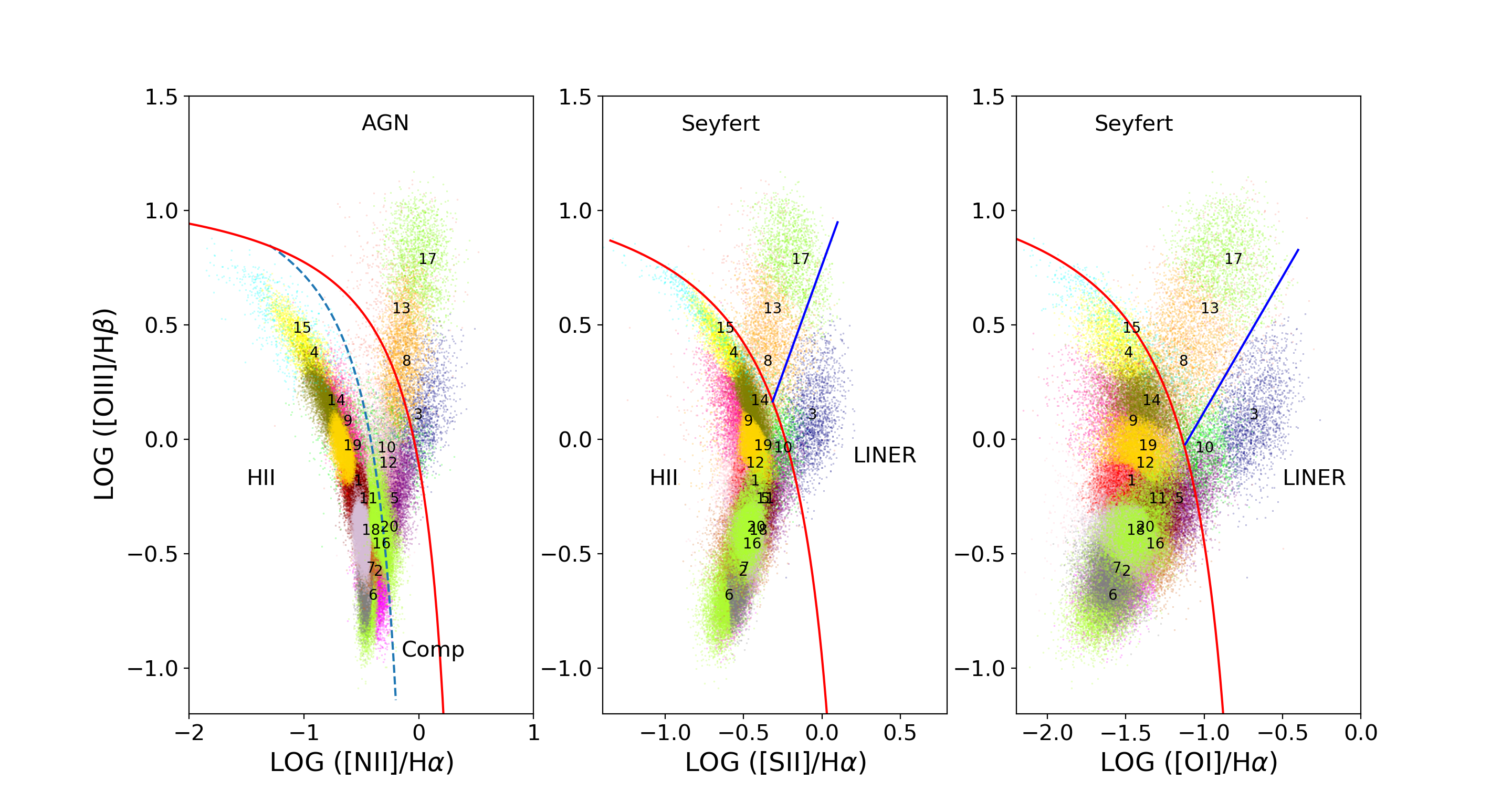}
\vspace{20pt}
\includegraphics[width=0.75\linewidth]{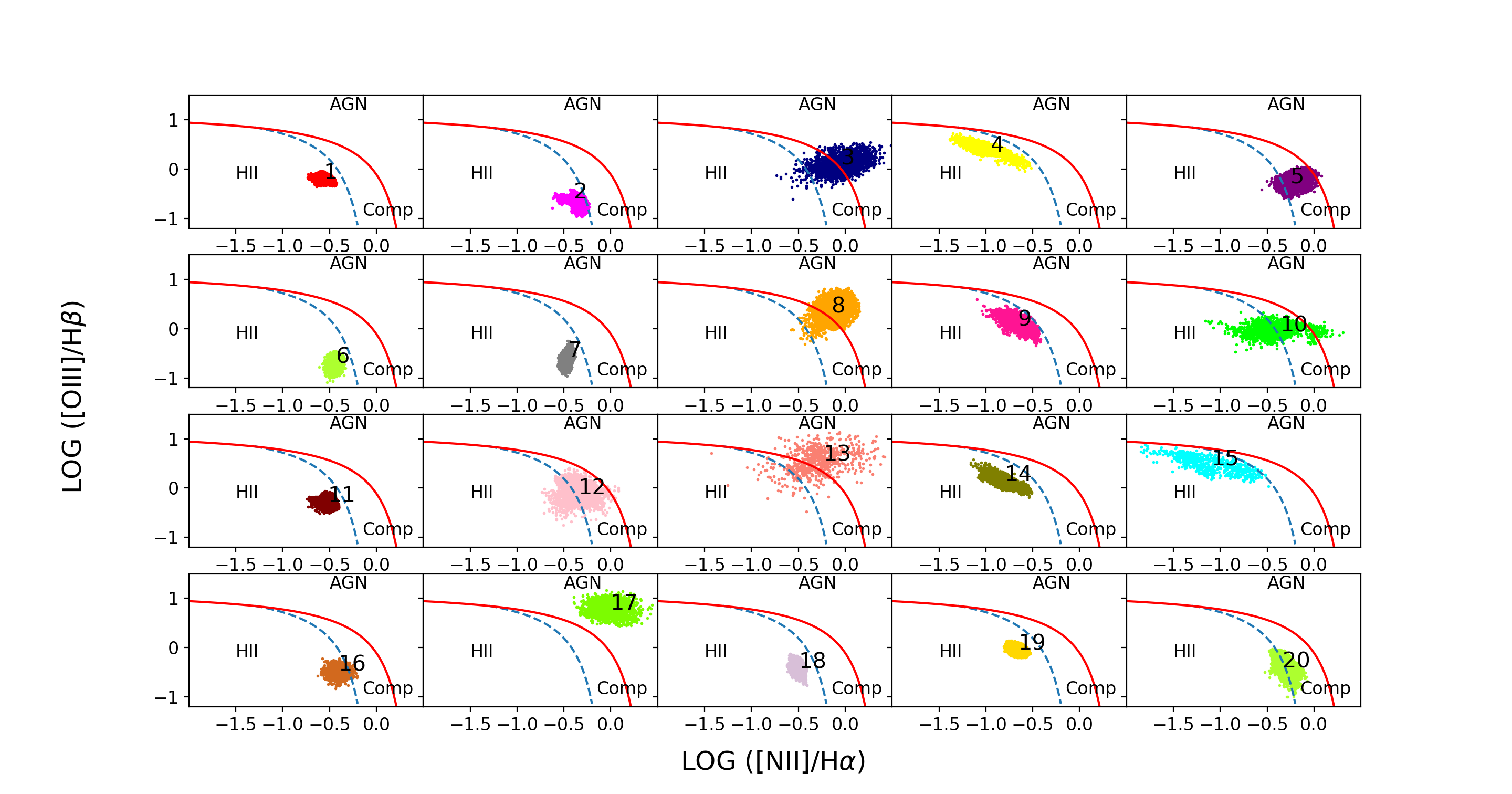}
\caption{The top panel shows the BPT diagnostic diagrams for the  SDSS DR8 sample with SNR$>$3 for all considered emission lines, including the  [\ion{O}{I}]$\lambda6300$. Each galaxy is colour-coded according to its most probable allocation to one of the 20 subpopulations. The maximum 'starburst' line of \citet{kewley2001theoretical} is shown  by the solid red line and the empirical upper bound on SFG of \citet{kauffmann2003host} is plotted as the dashed blue line. The empirical line for distinguishing  Seyferts and LINERs of  \citet{kewley2006host} is depicted by the solid blue line. The bottom panel shows the  20 subpopulations plotted on the [\ion{N}{II}]$/$H$\alpha$ vs [\ion{O}{III}]$/$H$\beta$ projection of the 4-dimensional diagnostic diagram. The subpopulations are numbered following the scheme in the top panel. This figure shows the spatial extent of each subpopulation and their location with respect to the standard diagnostic lines in the [\ion{O}{III}]$/$H$\beta$ diagram. Since these are 2-dimensional projections of the 4-dimensional distribution in each subpopulation, they only give an indication of the extent and location of each subpopulation.
}
\label{fig:20_OISNR}
\end{figure*}

{\newtwo{ We apply the SoDDA classifier in this new sample, by estimating the means, weights, and covariance matrices for the 20  sub-populations. We then assigned each sub-population to one of the 4 activity classes as presented in Table~\ref{tab:initial_comp_OISNR}. Figure \ref{fig:20_OISNR} shows the locations of the 20 sub-populations on the 2-dimensional projections of the diagnostic diagram. Comparison with Figs. \ref{fig:20},\ref{fig:comment7} shows very good agreement on the definition of the sub-populations in the two analyses, although the sub-populations in the new dataset (\ref{fig:20_OISNR}) appear to be more compact, as expected from the exclusion of the data with low  $[\ion{O}{I}]$\,$\mathrm{\lambda6300\AA}$ SNR.   Finally, we calculated the probability that each galaxy belongs to each one of the 4 classes (we will refer to the retuned SoDDA classifier as SoDDA-filtered).  We denote these probabilities, $\rho_{ic}^{\rm filter}$, to distinguish them from those computed with the "standard" sample used in \SS3, namely $\rho_{ic}$. 
}}

{\newtwo{A 3-way classification table that compares SoDDA-filtered with the  commonly used scheme proposed by \citet{kewley2006host} appears in Table~\ref{tab:3wayfiltered}. Each cell has 3 values: the number of galaxies with (i) $\rho_{ic}^{\rm filter}\geq 75\%$, (ii)   $50\%\leq \rho_{ic}^{\rm filter}<75\%$, and (iii)  $\rho_{ic}^{\rm filter}<50\%$, where $\rho_{ic}^{\rm filter}$ is the posterior probability that galaxy $i$ belongs to galaxy class $c$.  The results are in very good agreement with those presented in the original analysis (Table~\ref{tab:3way}). More specifically there is excellent agreement between the SoDDA-filtered and the \citet{kewley2006host} classification for the SFG, Seyfert, and LINER classes, in the sense that the fraction of objects in each class that have high-confidence ($\rho_{ic}^{\rm filter}\geq 75\%$) SoDDA classifications that agree with the \citet{kewley2006host} is either similar or larger in the case of the filtered sample. Only in the case of composite objects we have a slightly lower fraction of objects ($\sim1.5\%$) classified with SoDDA as such at high confidence. In addition while there was a considerable fraction of composite objects classified as such  at intermediate confidence ($50\%\leq\rho_{ic}<75\%$), in the analysis with the filtered sample this fraction is reduced, and there is an increased fraction classified as star-forming galaxies.  These small differences are the result of the slightly shifted means and slightly different widths of the sub-populations defined from the two samples, which are expected in a classifier that is trained on a subset of the sample. }} 

\begin{table}
\caption{The suggested classification of the {\new{20}} subpopulations means with SNR$>$3 for [\ion{O}{I}]$/H\alpha$}
\begin{tabular}{c r}
	Class & Subpopulation ID \\
	\hline
             SFG & {\new{1,  2,  4,  6,  7,  9, 11, 14, 15, 16, 18, 19, 20 }}\\
               Seyferts & {\new{8, 13, 17}}\\
                LINER & {\new{3}}\\
                Composites &{\new{5, 10, 12}} \\
\end{tabular}
\label{tab:initial_comp_OISNR}
\end{table}

\begin{landscape}
\begin{table}
\caption{ A 3-way classification table that compares the SoDDA  classification of the filtered sample with the standard, 2-dimensional classification scheme \citep{kewley2006host}. Each cell has 3 values: the number of galaxies with (i) $\rho_{ic}^{\rm filter}\geq 75\%$, (ii)   $50\%\leq \rho_{ic}^{\rm filter}<75\%$, and (iii)  $\rho_{ic}^{\rm filter}<50\%$, where $\rho_{ic}^{\rm filter}$ is the posterior probability that galaxy $i$ belongs to galaxy class $c$ under SoDDA filtered. {\VS Contradictory classifications are called ambiguous classifications by \citet{kewley2006host}}.}

\tabcolsep=0.11cm
\begin{tabular}{c c r r r r r r r r r r r r r r r r r r r  r r r r  }
\multirow{7}{*}{\rotatebox[origin=c]{90} {SoDDA filtered}}  & \multicolumn{24}{c}{Kewley et al. (2006) } \\
              & & \multicolumn{3}{c}{SFGs} &&  \multicolumn{3}{c}{Seyferts} &&  \multicolumn{3}{c}{LINERs} & & \multicolumn{3}{c}{Comp} &&  \multicolumn{3}{c}{Contradictory}  &&  \multicolumn{3}{c}{Total} \\

 &  & $\geq 75\%$ & $50\%-75\%$ & $<50\%$ &   & $\geq 75\%$ & $50\%-75\%$ & $<50\%$     & &      $\geq 75\%$ & $50\%-75\%$ & $<50\%$       &  &   $\geq 75\%$ & $50\%-75\%$ & $<50\%$      && $\geq 75\%$ & $50\%-75\%$ & $<50\%$       &&    $\geq 75\%$ & $50\%-75\%$ & $<50\%$  \\
 \cmidrule{3-5}\cmidrule{7-9}\cmidrule{11-13} \cmidrule{15-17} \cmidrule{19-21} \cmidrule{23-25} \\

 & SFGs & 73414 & 2338 & 25 &   & 0 & 0 &0     & &      0& 0 & 0       &  &   1973 & 1552 & 28      && 850 & 151 & 7       &&    76237 &  4041 &60  \\
&Seyferts            & 33 & 15 & 3 & &   3432 & 3 & 0    & &      41& 66 &11         &&   612 & 564 & 16      && 848 & 108 & 10       & &   4966 &  756 &40  \\
&LINERs          & 0 & 0 & 0&   & 0&0& 0      &&      965& 198 & 26         &&   466 &172& 16      && 43 & 39 & 5       &&    1474 &  409 &47 \\
&Comp     & 392 & 933 & 27 &     & 0 & 0 & 0 & &     3&45& 9         &&   4908 & 2719 & 49     & & 427 & 252 & 15      & &    5730 & 3949 &100  \\

\end{tabular}
\label{tab:3wayfiltered}
\end{table}
\end{landscape}

\begin{table*}
\caption{ Comparison of  the SoDDA classification of the filtered sample (SNR$>3$ on all considered diagnostic lines) with that of the SoDDA  classification of the sample considered in \SS3. The classification is performed in the  ([\ion{O}{III}]$/$H$\beta$,  [\ion{N}{II}]$/$H$\alpha$, [\ion{S}{II}]$/$H$\alpha$ and [\ion{O}{I}]$/$H$\alpha$) space.}
\begin{tabular}{c c c c c c c   }
\multirow{7}{*}{\rotatebox[origin=c]{90} {SoDDA filter}}  & \multicolumn{6}{c}{SoDDA } \\
              & & SFGs & Seyferts & LINERs & Composites   & Total \\
\cline{2-7}
 & SFGs & 78525 &     0      &       1         &   1812      & 80338  \\
&Seyferts            & 131 &    4751       &      45       &   835       &   5762  \\
&LINERs          & 18&    10      &      1757     &   145      & 1930 \\
&Composites     & 1698 &    12     &      110         &   7959    &  9779 \\
&Total    & 80372&   4773     &      1913        &   10751   &   \\

\end{tabular}
\label{tab:filtervsoriginal}
\end{table*}

\section{Online Material}

 In the online version of this article, we provide the following:
\begin{enumerate}
\item{Tables in numpy format including the estimated mean $\mu_k$, covariance matrix $\Sigma_k$, and the weight $\pi_k$ for each subpopulation $k=1,...,20$ (named \texttt{means.npy}, \texttt{covars.npy}, and \texttt{weights.npy} respectively). These are the definitions of the clusters as derived from the analysis presented is Section 3.}
\item{ {\newtwo{Tables in numpy format including the estimated mean $\mu_k$, covariance matrix $\Sigma_k$, and the weight $\pi_k$ for each subpopulation $k=1,...,20$ (named \texttt{m\_filter.npy}, \texttt{c\_filter.npy}, and \texttt{w\_filter.npy} respectively). These are the definitions of the clusters as derived from the analysis presented is Appendix A for the filtered sample.}}}
\item{Tables in numpy format providing the coefficients and the intercepts for the 4-dimensional (named \texttt{svm\_4d\_coefs.npy} and \texttt{svm\_4d\_intercept.npy})  and the 3-dimensional (named \texttt{svm\_3d\_coefs.npy} and \texttt{svm\_3d\_intercept.npy} respectively) surfaces based on the SVM method (Eqs. 12--23, and 24-35 respectively).  These are the definitions of the surfaces as derived from the analysis presented is Section 4. {\newtwo{We also include the trained SVM model, estimated using the {\tt scikit-learn} Python library, for both the 4-dimensional (\texttt{svm\_4d.sav}) and the 3-dimensional   (\texttt{svm\_3d.sav}) case.}} }
\item{A python script (\texttt{classification.py}) that allows the reader to directly apply the SoDDA and the SVM classification based {\new{on the clusters and the separating surfaces, respectively,  derived in Sections 3 and 4}}. It contains a function that given the 4 emission-line ratios $\log($[\ion{N}{II}]$/$H$\alpha$), $\log$([\ion{S}{II}]$/$H$\alpha$),  $\log$([\ion{O}{I}]$/$H$\alpha$) and $\log$([\ion{O}{III}]$/$H$\beta$), it computes the posterior probability of belonging to each of the 4 activity classes (SFGs, Seyferts, LINERs, and Composites {\newtwo{for class 0, 1, 2, and 3, respectively}}). We also include two functions which give the classification of a galaxy based on the 4-dimensional and the 3-dimensional SVM surfaces given its 4 emission line ratios.}
\item{ {\newtwo{A Readme file that explains the arguments and the output of the functions in the python script (\texttt{classification.py}) and contains examples of using them on sample data.}}}
\item{ {\newtwo{A table (\texttt{data\_classified.csv}) that contains the SoDDA-based probability that each galaxy  belongs to each one of the activity classes, derived in the analysis presented in Section 3. It also includes  the galaxy's SPECOBJID, the key diagnostic line-ratios, and the activity classification based on the class with the highest probability. }} }
\end{enumerate}


\bsp	
\label{lastpage}

\end{document}